\pdfoutput=1
\documentclass[prd,amsmath,floats,floatfix,
superscriptaddress,nofootinbib,showpacs]{iopart}

\usepackage[T1]{fontenc}
\usepackage[utf8]{inputenc}
\usepackage{lmodern}
\usepackage{float}

\usepackage{verbatim} 

\usepackage{appendix}

\usepackage[dvipsnames, usenames]{xcolor}
\definecolor{linkcolor}{rgb}{0.0,0.3,0.5}
\usepackage[hypertexnames=false, unicode, colorlinks=true, linkcolor=linkcolor,
citecolor=linkcolor, filecolor=linkcolor,urlcolor=linkcolor,
pdfusetitle]{hyperref}

\usepackage[all]{hypcap}
\usepackage{graphicx}
\usepackage{xspace}
\usepackage{amssymb}
\usepackage{makecell}
\usepackage[normalem]{ulem} 
\usepackage{bm} 
\usepackage{enumitem,amssymb}

\usepackage{microtype}

\usepackage[english]{babel}
\usepackage{blindtext}

\makeatletter
\newrobustcmd{\fixappendix}{%
  \patchcmd{\l@section}{1.5em}{7em}{}{}%
  \patchcmd{\l@subsection}{2.3em}{7em}{}{}%
}
\makeatother
\graphicspath{%
  {figs/}%
}

\DeclareMathAlphabet{\mathpzc}{OT1}{pzc}{m}{it}

\newlist{todolist}{itemize}{2}
\setlist[todolist]{label=$\square$}

\newcommand{\pare}[1]{\left(#1\right)}
\newcommand{\parea}[1]{\left[#1\right]}

\begin{document}

\title[]{Observational prospects of self-interacting scalar superradiance with next-generation gravitational-wave detectors}

\author{Spencer~Collaviti$^{1,2}$\footnote{Author to whom any correspondence should be addressed.}, Ling~Sun$^{1}$,  Marios~Galanis$^{3}$, Masha~Baryakhtar$^{4}$}
\address{$^{1}$OzGrav-ANU, Centre for Gravitational Astrophysics, College of Science, The Australian National University, Australian Capital Territory 2601, Australia}
\address{$^{2}$Institute of Physics, École Polytechnique Fédérale de Lausanne (EPFL), 1015 Lausanne, Switzerland}
\address{$^{3}$Perimeter Institute for Theoretical Physics, 31 Caroline Street North, Waterloo, Ontario, N2L 2Y5, Canada}
\address{$^{4}$Department of Physics, University of Washington, Seattle, Washington, 98195, USA}
\address{}
\address{E-mail: spencer.collaviti@epfl.ch}

\hypersetup{pdfauthor={Collaviti et al.}}

\date{\today}

\begin{abstract}
Current- and next-generation gravitational-wave observatories may reveal new, ultralight bosons. Through the superradiance process, these theoretical particle candidates can form clouds around astrophysical black holes and result in detectable gravitational-wave radiation. In the absence of detections, constraints---contingent on astrophysical assumptions---have been derived using LIGO-Virgo-KAGRA data on boson masses. However, the searches for ultralight scalars to date have not adequately considered self-interactions between particles. Self-interactions that significantly alter superradiance dynamics are generically present for many scalar models, including axion-like dark matter candidates and string axions. We implement the most complete treatment of particle self-interactions available to determine the gravitational-wave signatures expected from superradiant scalar clouds and revisit the constraints obtained in a past gravitational-wave search targeting the black hole in Cygnus X-1. We also project the reach of next-generation gravitational-wave observatories to scalar particle parameter space in the mass-coupling plane. We find that while proposed observatories have insufficient reach to self-interactions that can halt black hole spin-down, next-generation observatories are essential for expanding the search beyond gravitational parameter space and can reach a mass and interaction scale of $\sim 10^{-13}$--$10^{-12}$ eV$/c^2$ and $\gtrsim 10^{17}$~GeV, respectively. 

\end{abstract}

\tableofcontents
\markboth{}{}

\newpage

\section{Introduction}
\label{sec:introduction}

Massive bosonic particles whose Compton wavelength is comparable to the radius of black holes (BHs) can form bound states around them with exponentially growing occupation numbers~\cite{Ternov:1978gq,Detweiler,Zouros:1979iw,Dolan:2007mj}. This phenomenon is known as BH superradiance~\cite{Zel'Dovich1971}. The instability extracts energy and angular momentum from the BH until the energy density stored in the bound state backreacts on the spacetime~\cite{Misner:1972kx,Starobinsky:1973aij}. 

Astrophysical BHs are large enough that no Standard Model particle has an appropriate mass to induce the instability, but theories Beyond the Standard Model (BSM) generically predict the existence of many potential candidates. The most well-known example is the QCD axion, which solves the strong charge-parity (CP) problem \cite{Peccei1977,Weinberg1978,Wilczek1978} and is a leading dark matter candidate \cite{Preskill:1982cy,Abbott:1982af, Dine:1982ah,Marsh:2015xka,Adams:2022pbo}. Furthermore, string theory compactifications generically produce a multitude of light particles in the low-energy theory~\cite{green1987superstring,Svrcek_2006,Arvanitaki2010}. BH superradiance, first proposed in~\cite{Arvanitaki2010,Arvanitaki2011} as a probe of ultralight axions, is by now a well-established tool in the search for ultralight bosons~\cite{Brito:2014wla,Brito:2015oca} and an important particle physics target for current and future GW observatories~\cite{Sathyaprakash:2019yqt,Kalogera:2021bya,Baryakhtar:2022hbu,evans2023cosmic,Gupta:2023lga}. 

The principal signatures of BH superradiance are gravitational in nature: the spindown of astrophysical BHs~\cite{Arvanitaki2010,Arvanitaki2011,Witek:2012tr,East:2017ovw,Arvanitaki2015,Arvanitaki2017,East:2018glu,Ng:2019jsx}, emission of gravitational waves (GWs) \cite{Arvanitaki2010,Arvanitaki2011,Yoshino2014,Okawa:2014nda,Arvanitaki2015,Brito2017a,Brito2017b, Arvanitaki2017,East:2018glu,Siemonsen2023}, and additional phenomenology in binary systems~\cite{Zhang:2018kib,Hannuksela:2018izj,Zhang:2019eid,Baumann2019,Baumann:2019eav,Baumann:2021fkf}; see Ref.~\cite{Brito:2015oca} for further references. 
The BH is spun down due to the angular momentum extracted by the bound states. Thus, the observed BH spin in several astrophysical BHs has been used to constrain the existence of ultralight particles in the $\sim 10^{-13}$--$10^{-11}$ eV$/c^2$ range~\cite{Arvanitaki2015,Baryakhtar2021,Hoof:2024quk,Ng2021}.

Gravitational radiation is sourced by the time-dependent quadrupole of the ultralight particle cloud, which carries a large fraction of the initial BH energy. The emitted GWs are quasi-monochromatic and long-lived. This emission can potentially be reached by the current ground-based GW detector network, including Advanced LIGO (aLIGO)~\cite{Aasi_2015}, Advanced Virgo~\cite{aVirgo}, and KAGRA~\cite{KAGRA}, and future GW observatories, allowing for the direct search for new physics via GW observation.
Predictions for GW emission suggest that thousands of signals are expected from our galaxy alone if the particles exist in some parameter space~\cite{Arvanitaki2015,Arvanitaki2017,Brito2017a,Brito2017b,Zhu2020}. Furthermore, blind searches have been carried out for both continuous and stochastic GWs originating from scalar and vector ultralight particles~\cite{Abbot2022,Abbott2022MilkyWay,Palomba2019,Dergachev2019,Tsukada2019,Tsukada2021}. Non-detection of GWs in these blind searches can also be used to constrain ultralight bosons in the $\sim 10^{-13}$--$10^{-11}$ eV$/c^2$ range, but the constraints are sensitive to the mass and, especially, spin distributions of galactic BHs~\cite{Zhu2020}, which are poorly known. Directed searches targeting galactic BHs instead rely on the mass, spin, and age estimates of the targets~\cite{Sun2020}. Searches targeting remnant BHs from compact binary mergers are promising as the expected strength and form of the signals can be directly computed~\cite{Arvanitaki2017,Ghosh:2018gaw}, with the current and near-future ground-based GW detectors being sensitive enough to search for vectors in a range of masses~\cite{Baryakhtar:2017ngi,Isi2019,Jones2023}. 
Next-generation ground-based GW observatories, e.g., Cosmic Explorer~\cite{evans2021horizon,evans2023cosmic} and Einstein Telescope~\cite{Hild:2010id,Maggiore2020,Punturo:2010zz}, can reach above $\sim$~Gpc and tens of Gpc for scalars and vectors, respectively, in the $\sim10^{-14}$--$10^{-12}$~eV$/c^2$ mass range~\cite{Arvanitaki2017,Isi2019,Jones2023}.

While the superradiant instability only relies on gravity, interactions among the ultralight particles and possible couplings to the Standard Model or other sectors can drastically alter the evolution of the cloud of bound states~\cite{gruzinov2016black,Baryakhtar2021,Mathur:2020aqv,East:2022ppo,East:2022rsi,Siemonsen:2022ivj,Spieksma:2023vwl,Omiya2023,Omiya:2024xlz}. While these can be model-specific, both the QCD axion and the ultralight scalars predicted from higher-dimensional theories generically exhibit some level of self-interactions. 
Effects of self-interactions on scalar cloud evolution have been considered since early on~\cite{Arvanitaki2010,Yoshino2012,Yoshino2014}, but energy exchange processes between different bound states were largely neglected. Earlier, the belief was that strong attractive self-interactions lead to the collapse of the entire cloud back into the BH, a phenomenon dubbed a ``bosenova''~\cite{Arvanitaki2010}. However, as intuited first with a toy model by~\cite{gruzinov2016black} and later analyzed in detail for the full dynamical evolution of a BH-scalar system in~\cite{Baryakhtar2021}, self-interactions lead to energy exchange between different bound states, the BH, and infinity. Equilibrium is generally reached at bound-state occupations smaller than those required for the violent ``bosenova''. It is now well-established through both analytic analysis and numerical simulations~\cite{Baryakhtar2021,Omiya2023} that relatively strong self-interactions of the bosonic particles can lead to steady-states with orders of magnitude smaller occupation numbers than the expectation from purely gravitational growth. This prevents BH spin-down, keeps the cloud below the bosenova threshold, and suppresses GW emission.

It is therefore important to rigorously include the effects of self-interactions in searches for continuous GWs. In this work, we adopt the most up-to-date calculations of self-interacting scalars~\cite{Baryakhtar2021} to evolve the occupation numbers of self-interacting scalar clouds around stellar mass BHs, and compute the resulting quasi-monochromatic GW signal and its frequency drift.
We use the results to revisit the constraints on ultralight boson properties set by a past GW search targeting Cygnus X-1~\cite{Sun2020}.
Astrophysical measurements of the Cygnus X-1 accretion disk generally indicate its present spin is too high ($\geq 0.95$) \cite{Gou2011,Axelsson2011,Walton2016} to support a large boson cloud today (although see~\cite{Miller2009,Kawano2017}).
However, a superradiant cloud formed by particles with strong self-interactions would allow an old BH, like Cygnus X-1, to preserve its high spin. 
We also project the parameter space that could be explored in future searches targeting two example sources---Cygnus X-1 (an X-ray binary system) and MOA-2011 (a recently discovered, isolated galactic BH~\cite{Sahu2022})---with future terrestrial and space-based detectors: Cosmic Explorer~\cite{evans2021horizon,evans2023cosmic}, Einstein Telescope~\cite{Hild:2010id,Maggiore2020,Punturo:2010zz}, Matter-wave Atomic Gradiometer Interferometric Sensor-Space (MAGIS-Space)~\cite{Abe2021}, and Deci-hertz Interferometer Gravitational-wave Observatory (DECIGO)~\cite{Sato_2009}. We find that future observatories can probe non-negligible self-interactions in the mass range of $\sim 10^{-13}$--$10^{-12}$ eV/$c^2$.

This paper is organized as follows. In section \ref{sec:models}, we provide a brief review of superradiance in the presence of self-interactions. In section \ref{sec:calculation}, we present the expected GW signal and drift from example galactic BHs with a focus on Cygnus X-1. In section \ref{sec:results}, we update the constraints from a past directed search targeting Cygnus X-1 and project sensitivities for future observatories. We conclude and comment on future directions in section~\ref{sec:conclusion}. Finally, we provide details on the angular dependence of GW strains in~\ref{sec:Orientation Discussion}, explain the inference of the initial BH mass in~\ref{sec: Mass Evolution}, provide additional BH target examples in~\ref{sec:AdditionalResults} and present two qualitatively new, albeit minor, regimes of superradiant dynamics in~\ref{sec:minor_regimes}.


\section{Superradiance and self-interacting bosons}
\label{sec:models}

In this section, we review the theory and phenomenology of scalar field superradiance. We start with a brief summary of the superradiant instability (section~\ref{sec:SR}) and explain the particle theory of self-interactions (section~\ref{sec:self-interacting-bosons}). Next, we summarize the basic dynamics of self-interacting clouds (section~\ref{sec:SR-self-interacting-clouds}), focusing on the regimes of interest to the present work (section~\ref{sec:regimes}).

\subsection{The superradiant instability}
\label{sec:SR}

The superradiant instability occurs around rotating BHs and affects any wave whose frequency is lower than the angular frequency of the BH horizon. This is usually proven by showing that the flux at the horizon, as seen by an observer at infinity, can become \emph{negative}, showing that energy and angular momentum are extracted from the BH.

This flux is proportinal to the Killing vector tangent to the horizon. In Boyer-Lindquist coordinates this is $\xi=\partial_t+\Omega_H\partial_\phi$, where $\Omega_H \equiv \frac{c}{2r_g}\pare{\frac{\chi}{1+\sqrt{1-\chi^2}}}$ is the angular velocity of the horizon, with $\chi\equiv cJ/GM^2$ the dimensionless spin, $J$ and $M$ the angular momentum and mass of the BH, respectively, and $r_g\equiv GMc^{-2}$. A wave with angular frequency $\omega$ and (dimensionless) angular momentum $m$ about the spin axis of the BH has energy flux $\propto\omega(\omega-m\Omega_H)$ across the horizon, as measured by an observer at infinity. Therefore, if $\omega<m\Omega_H$, energy and angular momentum are extracted from the BH. 

Massive bosons with mass $m_b$, if light enough, form bound states around the BH, but the presence of the horizon induces imaginary parts to their frequencies \cite{Detweiler}. The states that satisfy the superradiant condition, $\omega<m\Omega_H$, will be unstable to growth. 
In the limit where the gravitational fine structure constant $\alpha\equiv m_b GM/(\hbar c)$ is much smaller than unity, the quasi-bound states are far from the horizon and are described by hydrogenic wavefunctions with the usual quantum numbers $(n,l,m)$, where $n$ is the principal quantum number, $l$ the total angular momentum, and $m$ the projection of the angular momentum of the state on the spin axis of the BH. The real parts of the bound state frequencies are determined by the energy levels~\cite{Detweiler,Baumann2019}
\begin{eqnarray}
    E_{nlm} = m_b c^2 \bigg(&1-\frac{\alpha^2}{2n^2}-\frac{\alpha^4}{8n^4}+\frac{(2l-3n+1)\alpha^4}{n^4(l+1/2)}\\
    &+\frac{2\chi m\alpha^5}{n^3l(l+1/2)(l+1)}\bigg)\nonumber.
    \label{eq:Energy Equation}
\end{eqnarray}
There are also corrections to the real parts of the frequencies due to attractive gravitational self-interactions of the cloud, which make it more bound. These can be computed by solving for the gravitational potential induced by the cloud through the Poisson equation, and then for the induced energy shifts to the energy eigenstates characterized by the appropriate $(n,l,m)$ numbers. Energy corrections from self-gravity depend on the occupation number of the cloud, and so are important for the calculation of frequency drifts. The computation was carried out in~\cite{Arvanitaki2015,Baryakhtar2021}. 
For simplicity, we use 211 to denote the $n=2,l=1,m=1$ level; similar for 322.
For the 211 and 322 levels, which we will see are the most significant ones in section~\ref{sec:SR-self-interacting-clouds}, these corrections are
\begin{eqnarray}
        \Delta  E_{211}&\approx -\frac{\alpha^3\hbar c^3}{G M^2}{\left(0.19 M_{211}+0.11M_{322}\right)},\\\Delta E_{322}&\approx -\frac{\alpha^3\hbar c^3}{G M^2}{\left(0.11 M_{211}+0.09M_{322}\right)},
\end{eqnarray}
where $M_{211}$ and $M_{322}$ are the masses of the respective level. 

The imaginary parts of bound state frequencies, $\omega_i$, are given by
\begin{equation}
    \omega_i\propto \alpha^{4l+5}\pare{m\Omega_H-E_n/\hbar}\pare{1+\mathcal{O}(\alpha)}.
    \label{eq:wi}
\end{equation}One can immediately deduce that gravitational superradiance will proceed hierarchically, as higher $l$ levels have much more suppressed growth rates due to their steeper angular momentum barriers.

\subsection{Self-interacting scalars}
\label{sec:self-interacting-bosons}

From a particle physics perspective, the simplest interaction of a real scalar field is a quartic self-interaction. Ultralight particles generally arise through non-perturbative physics at some very high scale, which produces an effective potential at low energies. Any potential of a field $\varphi$, expanded around a symmetric minimum, gives a mass term and a quartic self-interaction to the lowest order: 
\begin{equation}
    V(\varphi)\approx \frac{1}{2}\left(\frac{m_b c}{\hbar}\right)^2\varphi^2+\frac{\lambda}{24\hbar c}\varphi^4+\mathcal{O}(\varphi^6),
    \label{eq:potential}
\end{equation}
where $m_b$ is the scalar mass, and $\lambda$ is the dimensionless self-interaction coupling. It is, therefore, a quite typical prediction for ultralight particles to exhibit self-interactions.

The QCD axion is a well-known example of such a particle, generated at low energies as a result of a Peccei-Quinn (PQ) symmetry broken at a high energy scale $f_a$. At energies below the QCD scale, the axion acquires a symmetric potential. Because the low-energy axion potential is generated by QCD, its mass $m_b$ and the PQ-breaking scale $f_a$ are related by $m_bc^2\simeq 6\times 10^{-12}\,{\rm{eV}}(10^{18}\,{\rm{GeV}}/f_a)$,
resulting in quartic coupling uniquely fixed by the particle mass \cite{diCortona2016},
\begin{equation}
\lambda\simeq 0.3\frac{m_b^2c^4}{f_a^2}\simeq 10^{-80}\pare{\frac{m_bc^2}{10^{-12}\rm{eV}}}^4.
\end{equation}

More general axion-like particles (ALPs) can be present in the theory and are predicted to typically arise in string theory~\cite{Svrcek_2006,Arvanitaki2010}. While the symmetry-breaking scale $f$ and the particle mass $m_b$ are now related via unknown dynamics, a range of well-motivated scenarios can point to interesting targets such as string axions with $f\sim 10^{16}$-$10^{17}$~GeV, of order the grand-unification scale or string scale~\cite{Arvanitaki2010}. For convenience, we define the energy scale $f$ so that the dimensionless quartic coupling is $\lambda\equiv m_b^2 c^4/f^2$ and work in the $(m_b,f)$ parameter space.

If the axion-like particle comprises the dark matter and is produced via misalignment~\cite{Linde:1987bx,Turner:1990uz} with an $\mathcal{O}(1)$ initial angle and a time-independent potential, then the dark matter abundance is obtained for $f\simeq 3\times 10^{14}\,{\rm GeV}\,[10^{-12}\,{\rm eV}/(m_bc^2)]^{1/4}$, so that
\begin{equation}
\lambda\simeq \frac{m_b^2c^4}{f^2}\simeq 10^{-71}\pare{\frac{m_bc^2}{10^{-12}\,\rm{eV}}}^{5/2}.
\end{equation}
A range of initial misalignment angles, tuned to very small~\cite{Tegmark:2005dy} or very large~\cite{Arvanitaki2020} values,  predict that ALPs make up the dark matter abundance for $f$ in the range of $10^{13}$--$10^{16}$~GeV for axion masses relevant to superradiance of stellar-mass BHs. 
Parameter space of both the QCD axion and axion dark matter is probed by superradiance through the observation of old rapidly spinning BHs~\cite{Arvanitaki2015,Arvanitaki2017,Baryakhtar2021}, which has been used to set bounds on these particles in the $\sim 10^{-13}$--$10^{-11}$~eV$/c^2$ range.

\subsection{Superradiance of self-interacting scalar fields}
\label{sec:SR-self-interacting-clouds}

In this section we outline how the evolution of the scalar field $\varphi$ is modified in the presence of self-interactions. We focus on the processes that are relevant to this work and derive the corresponding equations describing the dynamics of the cloud. A more general treatment can be found in~\cite{Baryakhtar2021}.

A key simplification that allows for analytic computations is that the scalar field dynamics can be treated perturbatively for most of the parameter space. Perturbation theory applies as long as $(\hbar c)^{1/2}\varphi/f\ll1$, and in this regime a $\varphi^4$ term in the potential [Equation~(\ref{eq:potential})] accurately captures all the physics beyond the mass term.\footnote{Even when the potential has a slope and the first self-interaction coupling is $\propto\varphi^3$, it can be proven that the leading-order effects are equivalent to an effective quartic coupling~\cite{Baryakhtar2021}.} It can be proven that this is a good approximation for $\alpha\lesssim0.22$~\cite{Baryakhtar2021} initial conditions where 211 is the first level to grow, which is most of the relevant parameter space for GW observations in this work. We discuss breakdown of perturbation theory towards the end of section~\ref{sec:regimes}.

Writing the quartic term as $\mathcal{L}\supset \lambda \varphi^4/(24\hbar c)$, the equation of motion for the field is $(D^\mu D_\mu-m_b^2c^4)\varphi=-\lambda\varphi^3/(6\hbar c)$, where $D^\mu$ is the covariant derivative. In the non-relativistic limit, we can write $\varphi=\sqrt{2/m_b} \, {\rm{Re}}(\psi e^{-im_bc^2t/\hbar})$, where the variation timescale of $\psi$ is much longer than $m_b$. Then $\psi$ satisfies a Gross-Pitaevskii-like equation
\begin{equation}
\pare{i\partial_t+\mathcal{M}}\psi=-\frac{3\lambda}{24m_b^2c}\psi^2\psi^*,
\label{eq:eom_int}
\end{equation}
where $\mathcal{M}$ includes the usual non-relativistic Hamiltonian (kinetic energy, gravitational potential, etc.) and, crucially, the absorbing nature of the BH's horizon.
In deriving this equation, we have dropped terms oscillating at $3m_bc^2/\hbar$, which source relativistic radiation to infinity and are always subdominant.

Heuristically, one can think of Equation~(\ref{eq:eom_int}) as a perturbed Schr\"{o}dinger problem: given a background $\psi_c$ (for instance, the first level to grow), the right-hand side can be written at leading order as $|\psi|^2\psi^*\approx |\psi_c|^2\psi+\psi_c^2\psi^*$. The first term has the form of a time-dependent potential, with the potential $\propto|\psi_c|^2$. One can then compute energy shifts to all levels (eigenstates) in perturbation theory as in usual quantum mechanics. The complex conjugate term, $\psi_c^2\psi^*$, and the absorbing nature of the horizon make this problem more involved in practice, but not in essence. However, because both the $\psi_c^2\psi^*$ term and the absorbing horizon violate unitarity, they impart \emph{imaginary} frequency shifts, which can be positive or negative, thus leading to growth or decay of different levels. By energy conservation, these can all be understood as energy exchange processes between the different superradiant levels. We refer the reader interested in the details of this calculation to Ref.~\cite{Baryakhtar2021}.

\begin{figure}
\centering
\includegraphics[width=0.5\textwidth]{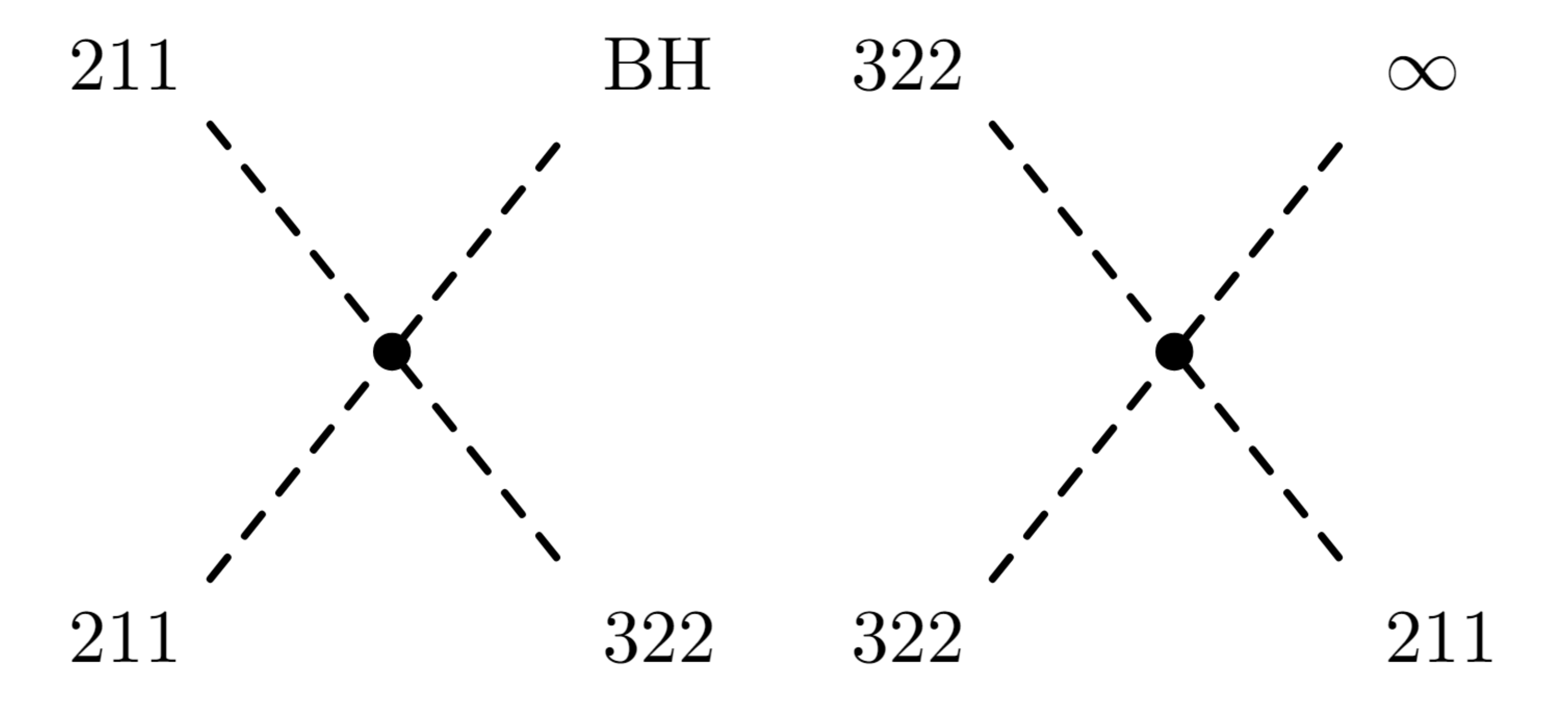}
\vspace{-0.4cm}
\caption[Diagrammatic representation of processes induced by self-interactions]{Diagrammatic representation of processes induced by self-interactions $211\times 211\to322\times {\rm BH}$ (left) and $322\times322\to 211\times\infty$ (right). The numbers represent the coupled $nlm$ levels, with the special characters `BH' and `$\infty$' representing particles in bound zero momentum states (which will be absorbed by the BH) and emitted to infinity (away from the BH), respectively.}
\label{fig:FeynmannSI}
\end{figure}

For $\alpha\lesssim 0.22$ and an initially growing 211 level, it can be proven that the cloud evolution is fully described by a closed two-level system of 211 and 322 \cite{Baryakhtar2021}: corrections to all other levels introduce growth (or decay) rates astrophysically irrelevant in this parameter space. The energy exchange between these levels stems from the terms $\psi^2_{211}\psi_{322}^*$ and $\psi^2_{322}\psi^*_{211}$. All other combinations shift the energies of the bound states and contribute to frequency drifts. Because of energy and angular momentum conservation, the oscillation sourced by the first term ($\psi^2_{211}\psi_{322}^*$) is bound and has zero angular momentum, while the latter ($\psi^2_{322}\psi^*_{211}$) is unbound. The first process pumps 322, depleting 211,\footnote{One can understand the sourcing of 322 via a bound mode that is decaying into the BH as a negative feedback process. More details can be found in Appendix A of Ref.~\cite{Baryakhtar2021}.} and the second process populates 211 by depleting 322 and emitting particle radiation to infinity. These terms in the evolution of non-relativistic field are depicted diagrammatically in Figure~\ref{fig:FeynmannSI}. The interplay between these processes and the superradiant growth of the cloud depends both on the gravitational binding parameter $\alpha$ and the self-interaction scale $f$.

In perturbation theory, where the spectrum does not get modified by self-interactions, the dynamics of the system can be fully characterized by the evolution of the occupation numbers. This is described by a set of coupled first-order ordinary differential equations that follow directly from energy and angular momentum conservation between the different levels, the absorbing BH and infinity, and were derived in~\cite{Baryakhtar2021}. The equations for the (normalized) occupation number of the levels, $\varepsilon_{nlm}\equiv N_{nlm}\hbar c/(GM^2)$, the dimensionless BH spin, $\chi$, and the mass of the BH, $M$ are:


\begin{eqnarray}
\dot{\varepsilon}_{211}=&\gamma_{\rm BH}^{211}\varepsilon_{211} -2\gamma^{\rm GW}_{211\times 211}\varepsilon_{211}^2 + \gamma_{322}^{211\times \rm GW}\varepsilon_{211}\varepsilon_{322} \nonumber\\
&-2\gamma^{322\times \rm BH}_{211\times 211}\varepsilon_{211}^2\varepsilon_{322}+\gamma_{322\times 322}^{211\times \infty}\varepsilon_{211}\varepsilon_{322}^2,\label{eq:SEE-1}\\
\dot{\varepsilon}_{322}=&\gamma_{\rm BH}^{322}\varepsilon_{322}-2\gamma_{322\times322}^{\rm GW}\varepsilon_{322}^2-\gamma_{322}^{211\times \rm GW}\varepsilon_{211}\varepsilon_{322}\nonumber\\
&+\gamma_{211\times 211}^{322\times \rm BH}\varepsilon_{211}^2\varepsilon_{322}-2\gamma_{322\times322}^{211\times \infty}\varepsilon_{211}\varepsilon_{322}^2,\label{eq:SEE-2}\\
\dot{\chi}&=-\gamma_{\rm BH}^{211}\varepsilon_{211}-2\gamma_{\rm BH}^{322}\varepsilon_{322}\label{eq:SEE-3},\\
\dot{M} &=\frac{ G M^2 m_b}{\hbar c}\left(-\gamma_{\rm BH}^{211}\varepsilon_{211}-\gamma_{\rm BH}^{322}\varepsilon_{322} + \gamma^{322\times \rm BH}_{211\times 211}\varepsilon_{211}^2\varepsilon_{322}\right).\label{eq:SEE-4}
\end{eqnarray}
We denote the rates in the form $\gamma_{\rm{initial\,state}}^{\rm{final\,state}}$. In particular, $\gamma_{\rm BH}^{nlm}$ describes the standard superradiant rate for the $nlm$ level and $\gamma_{nlm\times nlm}^{\rm GW}$ describes annihilations to GWs from the level $nlm$. Furthermore, $\gamma_{nlm\times nlm}^{n'l'm'\times\infty}$ describes the energy transfer of two quanta from the $nlm$ level to one quantum of the $n'l'm'$ level and to one quantum of radiation to infinity. $\gamma_{nlm\times nlm}^{n'l'm'\times{\rm BH}}$ describes a similar process where, instead, one quantum of energy falls back into the BH. As such, in Equations~(\ref{eq:SEE-1}) and (\ref{eq:SEE-2}), the first line corresponds to gravitational processes, namely superradiance and GW emission, while the second line corresponds to non-relativistic energy exchange processes between the different levels (see Figure~\ref{fig:FeynmannSI}). 

We solve these equations numerically using the Forward Euler method. Rates for the gravitational processes (superradiance and GW emission) are generated using a numerical package for superradiance, \texttt{Superrad} \cite{Siemonsen2023}, which natively includes relativistic corrections. Rates for the self-interaction processes are taken from~\cite{Baryakhtar2021}, where they are computed to the lowest order in $\lambda^2$. Relativistic corrections are then applied as described in Appendix B of~\cite{Baryakhtar2021}.

\begin{figure}
    \centering
    \includegraphics[width=1.0\textwidth]{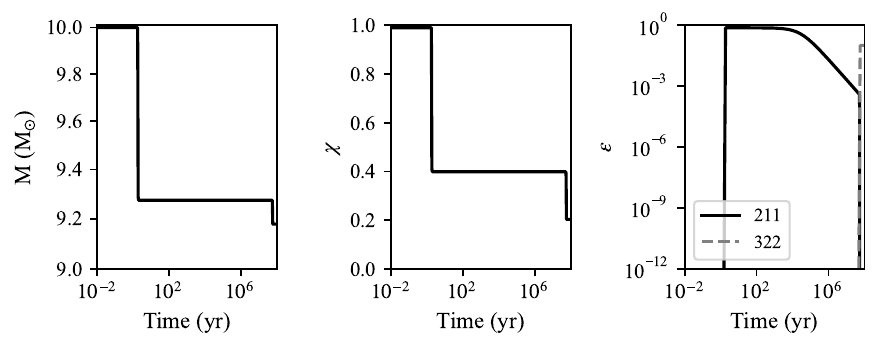}
    \includegraphics[width=1.0\textwidth]{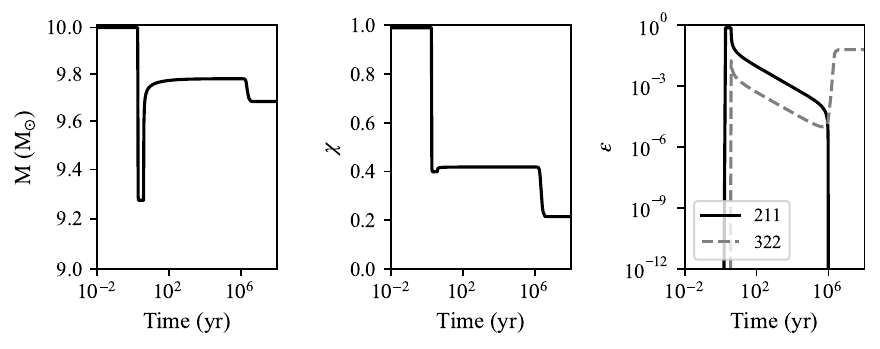}
    \caption{Evolution examples of (left) BH mass, (middle) BH spin, and (right) ultralight boson level occupations. The boson mass is $m_b=1.5\times10^{-12}$ eV/c$^2$ and the BH has an initial mass of $M_i = 10 M_\odot$ and an initial spin of $\chi_i=0.99$. The self-interaction parameter is $f=10^{19}$~GeV for the top panels (gravitational regime) and $f=10^{17}$~GeV for the bottom panels (moderate self-interaction regime).}
    \label{fig:Combined Regimes}
\end{figure}

\subsection{Regimes of scalar superradiance}
\label{sec:regimes}

Here we outline the behaviour of self-interacting superradiant clouds in the $(m_b,f)$ plane. Four main regimes are described in Ref.~\cite{Baryakhtar2021}, in the order of increasing self-interaction strength: the gravitational regime, the moderate self-interaction regime, the strong self-interaction regime, and the no spin-down regime.
The first two, the gravitational and moderate self-interaction regimes, are demonstrated in this paper to be of particular observational interest. Example evolutions of systems in these regimes are shown in the upper and lower panels of Figure~\ref{fig:Combined Regimes}, respectively, and we here focus on the distinct dynamics and signatures of only these two regimes.
A detailed analysis of all four regimes can be found in Ref.~\cite{Baryakhtar2021}. 

\emph{The gravitational regime (purely-gravitational growth):} The growth of the bound states starts as soon as a fast-spinning BH is born, such as immediately after a merger or a supernova. If the BH is spinning fast enough, 211 is the fastest-growing level, followed by 322. The superradiance rates for higher levels are generally relevant on a timescale longer than the age of the target BHs, so only the 211 and 322 levels are important. 

By energy and angular momentum conservation, the exponentially growing quasi-bound state extracts energy and angular momentum from the BH. The BH spin is primarily lost due to angular momentum extraction, with the mass extraction a factor of $\alpha/l$ subdominant. Because astrophysical BHs have high angular momenta
\begin{equation}
J=\chi GM^2c^{-1}\simeq 10^{78}\hbar\chi\pare{\frac{M}{10M_\odot}}^2,
\end{equation}
it takes roughly $\log(J/(m\hbar))\sim180$ e-folds of superradiant growth to backreact on the spacetime. As a result, in the steady-state configuration, the BH will have been spun down by an $\mathcal{O}(1)$ fraction, and the cloud will have a high occupation number $\sim \mathcal{O}(J/\hbar)$. This is seen in the upper panels of Figure~\ref{fig:Combined Regimes} at a time of $\sim 10^{0}$ yr. To leading order in $\alpha$, the final BH spin after saturation of the 211 level is $\chi\approx \chi_i-4\alpha/(1+4\alpha^2)$, where $\chi_i$ is the initial spin of the BH.
The energy density stored in a cloud around a stellar-mass BH can be as large as that of a white dwarf, $\sim 10^9$~kg/m$^3$, so any pre-existing scalar abundance in the neighborhood of the BH does not appreciably change either the final occupation number or the required e-folding time.

The bosonic cloud has an oscillating quadrupole moment, which sources gravitational radiation. The emission occurs at an angular frequency of $\sim 2m_bc^2/\hbar$, which can heuristically be thought of as boson annihilations. Because of the non-relativistic nature of the hydrogenic-like clouds, GW power $P$ is a steep function of $\alpha$, $P\propto GN^2 m_b^4\alpha^{4l+16}\hbar^2c^3$, where $N$ is the occupation number of the cloud~\cite{Yoshino2014}. Nevertheless, the decay timescale can be short compared to astrophysical timescales for stellar-mass BHs, which manifests as cloud depletion over time.

The next level to grow will reach its maximum occupation after $\sim 180$ e-folds of its superradiant growth. As the BH is spun down to an even smaller spin, the previous level becomes unstable to decay and gets re-absorbed. This is seen in the upper panels of Figure~\ref{fig:Combined Regimes} at a time of $\sim 10^{8}$ yr. Because of the hierarchy of superradiant timescales, the previous level has been depleted appreciably due to GW emission when it falls back into the BH. The angular momentum of the black hole continues to decrease and the mass can temporarily increase as the now unstable level gets absorbed. Thus the evolution proceeds in the same spirit, until astrophysical timescales become relevant, such as the BH's age or accretion rate.

The gravitational regime of the mass-interaction parameter space encapsulates the interaction strengths of QCD axions with masses relevant for superradiance~\cite{Baryakhtar2021}. GW annihilation signatures are the loudest in this regime.

\emph{The moderate self-interaction regime:} The crossover from the gravitational regime is characterized by two conditions: 1) the age of the BH, $T_{\rm{BH}}$, is large enough to allow the process $211\times 211\to322\times {\rm BH}$ to appreciably populate 322, and 2) the first level, 211, has not been significantly depleted by GW emission by this time. This computation has been carried out in Section IV.B.3 of~\cite{Baryakhtar2021}. Schematically, the first condition amounts to using Equation~(\ref{eq:SEE-2}) and demanding that the self-interactions-induced process $\gamma_{211\times 211}^{322\times{\rm BH}}(\varepsilon_{211}^{\max})^2\gtrsim \log(GM^2)/T_{\rm BH}$, namely the time it takes to build up 322 in the lifetime $T_{\rm BH}$ of the BH. The second condition similarly comes about from Equation~(\ref{eq:SEE-1}) and is the requirement $\gamma_{211\times 211}^{\rm GW}\varepsilon_{211}\lesssim \gamma_{211\times 211}^{322\times{\rm BH}}(\varepsilon_{211}^{\max})^2/\log(GM^2)$. Combining these two conditions, the moderate self-interaction regime occurs for
\begin{eqnarray}
f<f_1\approx\min\bigg[&3\times 10^{16}\,\textrm{GeV}\pare{\frac{T_{\rm{BH}}}{10^{10}\,\textrm{yr}}}^{1/4}\pare{\frac{m_bc^2}{10^{-13}\,\textrm{eV}}}^{1/4}\pare{\frac{\alpha}{0.01}}^{11/4},\nonumber\\
&8\times 10^{18}\,\textrm{GeV}\pare{\frac{0.01}{\alpha}}^{3/4}\pare{\frac{\chi}{0.9}}^{1/4}\bigg].
\end{eqnarray}

A representative time evolution of the occupation numbers and BH mass and spin is shown in the lower panels of Figure~\ref{fig:Combined Regimes}. Here, the 211 level grows unimpeded to its maximum occupation number, $\varepsilon_{211}\approx 1$, and spins down the BH as in the purely gravitational evolution. Because of the stronger self-interactions, the 322 level can grow much faster than $\sim 180$ e-folds of its superradiant growth. 
As the 322 level grows to an appreciable size, re-pumping of the 211 level through $322\times322\to 211\times\infty$ becomes significant and the cloud enters a regime where the ratio $\varepsilon_{211}/\varepsilon_{322}>1$ remains constant, forming a quasi-equilibrium. This quasi-equilibrium can be seen in the lower panels of Figure~\ref{fig:Combined Regimes} for times between $\sim10^0$ yr and $10^6$ yr. The two-level cloud loses energy to the BH and to infinity, primarily through quartic processes. 

Since the 322 level grows through self-interactions to a very large occupation, its superradiant growth drives it to maximum occupation in just a few e-folds, thus reaching saturation much earlier than in the purely gravitational growth regime. As shown in the rightmost panels of Figure~\ref{fig:Combined Regimes}, for moderate self-interactions, 322 has grown maximally in $\sim10^6$~yr (lower panels) compared to $\sim 10^8$~yr in the purely-gravitational growth regime (upper panels). 

Once the quartic process $211\times211\to 322\times{\rm BH}$ becomes faster than the superradiance of 211, the initial growth of 211 is suppressed by a simultaneous growth of 322. Quantitatively, the condition here comes about from Equation~(\ref{eq:SEE-1}) by demanding $\gamma_{211\times 211}^{322\times{\rm BH}}(\varepsilon_{211}^{\max})^2\gtrsim 2\log (GM^2)\gamma_{\rm BH}^{211}$. However, for small $\alpha$, the system equilibrates first, before this condition can be met. Setting $\dot{\varepsilon}_{211}=\dot{\varepsilon}_{322}=0$ in Equations~(\ref{eq:SEE-1}) and~(\ref{eq:SEE-2}) gives a different scaling for the threshold $f$ to transition to the moderate self-interaction regime for $\alpha\lesssim 0.04$. We refer the reader to Section IV.B.4 and Appendix E of~\cite{Baryakhtar2021} for the detailed computation. In both cases, we have a much smaller maximum occupation of the 211 level, $\varepsilon_{211}^{\max}\ll1$, which occurs for
\begin{equation}
f_1>f>f_2\approx2\times 10^{16}\textrm{GeV}\pare{\frac{\chi(t_0)}{0.9}}^{1/4}\min\parea{\pare{\frac{\alpha}{0.04}}^{3/4},\pare{\frac{\alpha}{0.04}}^{3/2}}.
\end{equation}
When $f$ approaches $f_2$, it marks the boundary between the moderate self-interaction regime and strong self-interaction regimes.

In the moderate interaction regime, GW emission from 211 annihilations can be strong enough to be detected, but the signal duration is shortened. A novel gravitational signature here is GWs through level transitions, which are rare and short-lived in the gravitational regime (see~\cite{Arvanitaki2015,Siemonsen:2019ebd}).  These $322\to 211+ {\rm GW}$ transition signals have angular frequencies $\sim \alpha^2m_bc^2/\hbar\ll m_bc^2/\hbar$, so can be relevant to future deci-hertz GW detectors. For $f<f_2$, the cloud occupation numbers are strongly suppressed, and the emitted GWs do not have observational prospects (see Section~\ref{sec:results}).

\emph{Breakdown of perturbation theory:} In all of the aforementioned regimes, the perturbative treatment of self-interactions is justified in the sense that the hydrogenic wavefunctions are not significantly distorted, timescales for all rates are much longer than oscillation times, and the dynamics are restricted to a two-level system. For $\alpha\gtrsim 0.22$, the two-level approximation no longer holds and higher levels can grow. In this case, additional signatures and transitions between higher levels can be present~\cite{Baryakhtar2021,Omiya:2024xlz}. While perturbative evolution can, in principle, be hierarchical and closed in non-gravitational regimes for $\alpha\lesssim 0.3$~\cite{Baryakhtar2021}, no demonstration of a closed regime which is under good analytic control has been performed for $\alpha\gtrsim 0.3$. We do not include these high-$\alpha$ scenarios in our current work as we only have control of our approximations in the purely gravitational growth regime, where $f$ is close to the Planck scale. There are potential signatures in the regime of $\alpha\gtrsim 0.22$, as shown in the results of this study, so we stress the importance of studying the high-$\alpha$ dynamics and parameter space in the future.

\section{Gravitational wave signatures}
\label{sec:calculation}

In this section, we select two galactic BHs as potential targets for GW searches (Section~\ref{subsec:targets}), compute the boson occupation numbers and GW strain as a function of boson mass and self-interaction strength (Section~\ref{subsec:occupation_strain}), and discuss the frequency drift of the GW signal (Section~\ref{subsec:fdot}).

\subsection{Galactic targets for continuous wave searches}
\label{subsec:targets} 
Known galactic BHs (usually with reasonably well-measured mass and, in some cases, spin) are closer and better localized in the sky compared to remnant BHs formed in binary mergers and thus are promising targets if they host boson clouds through superradiance~\cite{Isi2019}.
Here, we consider two known galactic BHs of masses on the order of $10 M_\odot$: Cygnus X-1 and MOA-2011. Cygnus X-1, a BH in an X-ray binary, has been previously studied as a potential host for ultralight scalars in a directed search~\cite{Sun2020}.  We update the search results obtained in Ref.~\cite{Sun2020} using the model investigated in this paper, still assuming a high initial spin and a final spin determined by superradiant evolution. MOA-2011 is taken as an example of isolated BH targets, which do not suffer sensitivity loss due to the spread of signal power caused by the binary motion: the binary orbital parameters are typically not measured well enough to fully account for the Doppler frequency modulation in continuous wave searches~\cite{Isi2019,Riles2023}.
Another potential target is the 33 $M_\odot$ BH in a long-period orbit (11.6 yr) with a companion star of 0.76 $M_\odot$ at a distance of 590~pc recently observed by Gaia~\cite{GaiaBH3}. The long orbital period and small semimajor axis make the BH a potentially interesting target since a search would suffer less from the Doppler modulation effect. We do not include this BH observed by Gaia in this study, but BHs of a similar kind may be of interest in future work. 

The parameters we use for the two example BHs, Cygnus X-1 and MOA-2011, are summarised in Tables~\ref{tab:Cygnus parameters} and \ref{tab:MOA parameters}. 
For Cygnus X-1, we adopt the same source parameters as used in Ref.~\cite{Sun2020} and consider two assumptions of the BH age, a younger estimate at $1.0 \times 10^5$ yr and an older estimate at $1.0 \times 10^6$ yr (slightly different from those in Ref.~\cite{Sun2020}) to derive the updated constraints from the existing search for self-interacting scalars.
For MOA-2011, we use the known parameters available in literature~\cite{Sahu2022}, assume an optimal initial spin of $0.99$ (as it is unknown), consider a range of BH ages, and consider the most-probable, edge-on orientation (inclination, $\iota=90^{\circ}$). The dependence of the GW strain on inclination is discussed in \ref{sec:Orientation Discussion}. For both BHs, we consider ultralight scalar masses corresponding to a range of $\alpha \in [0.01, 0.20]$.

We note that although the mass and distance parameters adopted for Cygnus X-1 are consistent with the estimates by Orosz \textit{et al.}~\cite{Orosz2011}, Cabellero-Nieves \textit{et al.}~\cite{Caballero-Nieves2009}, and others, they differ from the recent result by Miller-Jones \textit{et al.}~\cite{Miller-Jones2021}. In particular, a larger BH mass for Cygnus X-1 ($M=21.2\pm0.2 M_\odot$) is proposed in Ref.~\cite{Miller-Jones2021} than that considered previously ($M=14.8\pm1.0M_\odot$). This could affect the reachable parameter space, since the expected signal strength increases with BH mass. Such a larger BH mass would also shift the optimal boson mass (i.e. the one producing the strongest signal) to a mass approximately 25\% smaller. It may be worth carrying out a new search using parameters presented in Ref.~\cite{Miller-Jones2021} with more sensitive detector data.

When calculating the cloud evolution for these BH-boson systems, since the BH mass evolves due to the cloud growth, and the measured BH mass is the final state today, it is necessary to estimate the initial BH mass. The retrospective procedure to determine the initial BH mass is described in \ref{sec: Mass Evolution}.

\begin{table}
    \centering
    \begin{tabular}{ |c c c c| } 
     \hline
     Property & Symbol & Value & Reference \\ 
     \hline
     Mass ($M_{\odot}$) & $M$ & 14.8 & \cite{Orosz2011} \\ 
    Assumed Initial Spin & $\chi$ & 0.99 & \cite{Gou2011} \\ 
     Inclination & $\iota$ & $27.1^\circ$ & \cite{Orosz2011} \\
     Distance (kpc) & $d$ & 1.86 & \cite{Reid2011} \\
    Age -- Old Estimate (yr) & $T_{\rm bh}$ & $1.0\times 10^6$ & \cite{Gou2011,Wong2012} \\
     Age -- Young Estimate (yr) & $T_{\rm bh}$ & $1.0\times 10^5$ & \cite{Sun2020,Russell2007} \\
     \hline
    \end{tabular}
    \caption{Cygnus X-1 benchmark parameters used in the reach calculations. The spin today is measured to be high and is unlikely to have grown through accretion~\cite{Gou2011}.}
    \label{tab:Cygnus parameters}
\end{table}

\begin{table}
    \centering
    \begin{tabular}{ |c c c c| } 
     \hline
     Property & Symbol & Value & Reference \\ 
     \hline
     Mass ($M_{\odot}$) & $M$ & 7.3 & \cite{Sahu2022} \\
     Assumed Initial Spin & $\chi$ & 0.99 & - \\
     Assumed Inclination & $\iota$ & $90.0^\circ$ & -\\
     Distance (kpc) & $d$ & 1.58 & \cite{Sahu2022} \\
     \hline
    \end{tabular}
    \caption{MOA-2011 benchmark parameters used in the reach calculations. No measurements are available for the initial BH spin and inclination, and the assumed values used in this study are listed.}
    \label{tab:MOA parameters}
\end{table}

For brevity, we present only results specific to Cygnus X-1 with an assumed age of $10^5$ years in this section. Results for Cygnus X-1 with an assumed older age of $10^6$ years and for MOA-2011 are obtained in the same way and are presented in \ref{sec:AdditionalResults}.

\subsection{Occupation numbers and gravitational-wave strain}
\label{subsec:occupation_strain}
\begin{figure}
    \centering
    \includegraphics[width=1.0\textwidth]{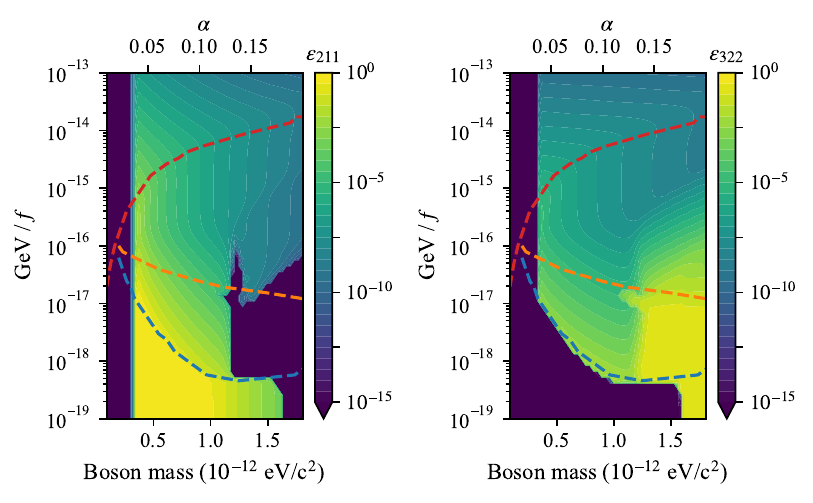}
    \caption{Normalised occupation numbers, $\varepsilon_{211}$ (left) and $\varepsilon_{322}$ (right), for the 211 and 322 levels in the boson cloud around Cygnus X-1, as a function of boson mass ($m_b$) and inverse interaction parameter ($1/f$), assuming parameters as in Table~\ref{tab:Cygnus parameters} and a BH age of $1.0\times10^5$~yr, if the corresponding boson exists. Dashed curves divide the entire parameter space into four regimes, from top to bottom: the no spin-down regime, strong self-interaction regime, moderate self-interaction regime, and gravitational regime.}
    \label{fig:Cygnus 1e5 yr occupation}
\end{figure}

We first consider the normalized occupation numbers of the 211 and 322 levels, $\varepsilon_{211}$ and $\varepsilon_{322}$, for Cygnus X-1 assuming an age of $10^5$ years (calculated using the model and numerical integration method described in Section~\ref{sec:models}). These occupation numbers are shown as coloured contours in Figure~\ref{fig:Cygnus 1e5 yr occupation}. 
The four regimes of ultralight boson self-interaction: the no spin-down regime, strong self-interaction regime, moderate self-interaction regime, and gravitational regime, are separated from top to bottom by the red, orange, and blue dashed curves.

As the inverse of the interaction parameter, $1/f$, increases (equivalently, the strength of ultralight boson self-interactions increases), the occupation numbers of the 211 and 322 levels, $\varepsilon_{211}$ and $\varepsilon_{322}$, generally decrease. 
This is consistent with the fact that self-interactions lead to energy loss from the BH-boson system via the emission of particles to infinity. 
As the boson mass increases, we see a general trend of $\varepsilon_{211}$ decreasing and $\varepsilon_{322}$ increasing in the gravitational and moderate self-interaction regimes. Such trend in $\varepsilon_{211}$ and $\varepsilon_{322}$ as a function of $m_b$ can be explained as heavier bosons lead to faster cloud evolution  (as the superradiance and self-interaction rates are both higher). 
For the 211 level, which always grows first and then decays, faster evolution means a lower occupation number is expected at a given age, e.g., $10^5$~yr considered here. For the 322 level, which grows slower, faster evolution can lead to a greater occupation.
For $m_b \gtrsim 1.2 \times 10^{-12}$~eV/c$^2$ and $m_b \gtrsim 1.6 \times 10^{-12}$~eV/c$^2$ in the moderate self-interaction regime and gravitational regime, respectively, the 211 level has decayed significantly so $\varepsilon_{211}$ drops to a minimal level (dark bottom right corner in the left panel), while $\varepsilon_{322}$ becomes dominant (bright bottom right corner in the right panel) at the chosen age.
Similarly, the dark region in the right panel for lighter bosons in the gravitational regime corresponds to the parameter space where the 211 level occupation is dominant, while the 322 level has not grown to an appreciable level. The dark regions on the left-hand side of the left panel corresponds to the parameter space where the 211 level, although dominant, has not reached its maximum size within the BH's age.

More complicated behaviour is seen in two additional, exceptional regimes which occur in small parts of the parameter space and are discussed in \ref{sec:minor_regimes}. The first, the `immediate (level) switch' regime, occurs around $m_b\sim1.1\times10^{-12}$~eV/c$^2$ and GeV/$f\sim1.5\times10^{-17}$ (for the particular BH example considered here; see Figure~\ref{fig:Cygnus 1e5 yr occupation}). In this regime, low $\varepsilon_{211}$ and high $\varepsilon_{322}$ occupations are observed due to an immediate switch from 211 dominant to 322 dominant in the BH-boson system. The second, the `harmonic equilibrium' regime, occurs for $m_b \gtrsim 1.2 \times 10^{-12}$~eV/c$^2$ and GeV/$f \sim 10^{-17}$--$10^{-16}$. In this regime, although the 322 level is dominant, the 211 level persists due to a harmonic equilibrium driven by self-interactive couplings.

\begin{figure}
    \centering
    \includegraphics[width=1.0\textwidth]{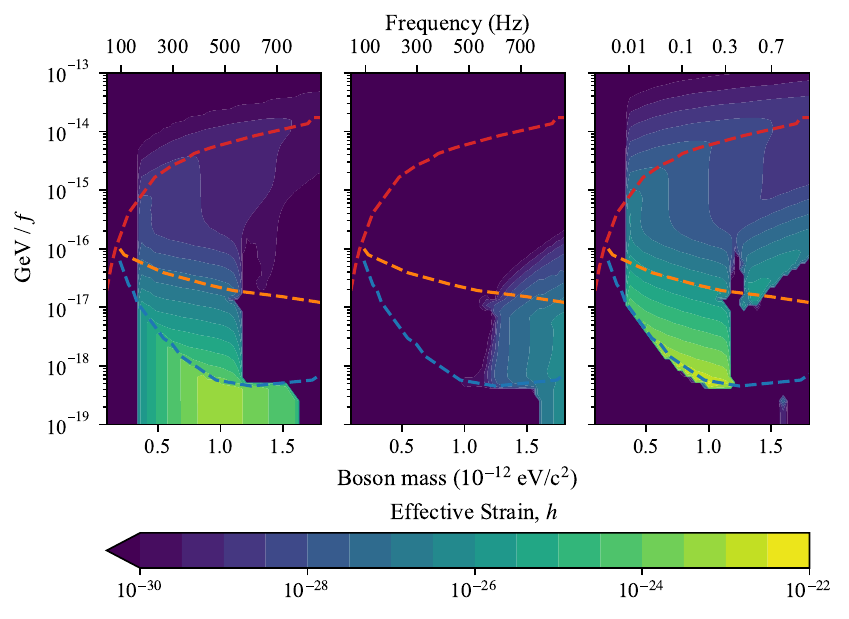}
    \caption{GW strain, $h$, of the boson signal from Cygnus X-1, assuming BH parameters as in Table~\ref{tab:Cygnus parameters} and an age of $1.0\times10^5$ yr. The left, middle, and right panels show $h$ corresponding to $211$ annihilation, $322$ annihilation, and $322\rightarrow211$ transition, respectively. Dashed curves divide the parameter space into four regimes, from top to bottom: the no spin-down regime, strong self-interaction regime, moderate self-interaction regime, and gravitational regime.}
    \label{fig:Cygnus 1e5 yr strain}
\end{figure}

Next, we estimate the effective GW strain, denoted by $h$, which we would measure on Earth depending on the orientation of the BH-boson system, for the three emission mechanisms. See \ref{sec:Orientation Discussion} for the angular dependence.
In Figure~\ref{fig:Cygnus 1e5 yr strain}, we show in colored contours the expected GW strain for signals from Cygnus X-1 with an age of $10^5$ years. From left to right, we show the strain from the 211 annihilation, 322 annihilation, and $322\rightarrow 211$ transition modes, respectively, as a function of boson mass (or corresponding emission frequency on the top axis) and inverse interaction parameter ($1/f$). 

The trends in GW strain over the parameter space, as shown in Figure~\ref{fig:Cygnus 1e5 yr strain}, can be rationalised in terms of three factors. The first factor is the occupation number of the relevant state(s). For a given boson mass (or, equivalently, $\alpha$), the GW strain is proportional to the two relevant normalised occupation number(s)~\cite{Baryakhtar2021}. That is, the strain from 211 annihilation, 322 annihilation, and $322\rightarrow211$ transition is proportional to $\varepsilon_{211}^2$, $\varepsilon_{322}^2$, and $\varepsilon_{211}\varepsilon_{322}$, respectively. 
Thus, for each GW generation mechanism, when there is a low occupation in the relevant state(s), the strain is weak and the region appears dark (i.e., the GW signal strength is minimal). Thus, the dark regions in the left ($\varepsilon_{211}$) and right ($\varepsilon_{322}$) panels of Figure~\ref{fig:Cygnus 1e5 yr occupation}  correspond to dark regions in the left (211 annihilation) and middle (322 annihilation) panels of Figure~\ref{fig:Cygnus 1e5 yr strain}. When either the 211 or 322 region in Figure~\ref{fig:Cygnus 1e5 yr occupation} is dark, there is a dark region for the transition signal in Figure~\ref{fig:Cygnus 1e5 yr strain}. The second factor is the boson mass, or equivalently, $\alpha$. 
In general, a higher boson mass (larger $\alpha$) leads to stronger GW emission (brighter in the figure), provided that the state(s) are sufficiently occupied. 

Finally, the third factor is the hierarchy of GW strengths for the different generation mechanisms. For the same boson mass, when $\varepsilon_{211} \sim \varepsilon_{322}$, $322\rightarrow211$ transition signals are much stronger than 211 annihilation signals, which are, in turn, much stronger than 322 annihilation signals. The relative enhancement of the transition signal compared to the annihilation signal, for \emph{fixed} cloud occupation numbers, can be qualitatively understood from the length scales associated with GW emission. Radiation is sourced by the bound state whose typical size is the gravitational Bohr radius, $\sim \hbar/(\alpha m_b c)$, while the wavelength of the radiation is $\sim\hbar/(2 m_b c)$ for annihilations and $\sim72\hbar/(5\alpha m_b c)$ for transitions. Therefore, there is a scale mismatch for annihilation radiation where a large source is emitting a high-frequency signal, leading to a cancellation of power, which is even more significant for higher angular momentum clouds (322 vs 211). In contrast, transition radiation has a longer wavelength compared to the source size and so its power is a less steep function of $\alpha$. Nevertheless, transitions occur only in the moderate self-interaction regime where occupation numbers are suppressed compared to their maximum. This can cancel some of this enhancement and yield a strain of comparable or even smaller size than that from annihilations in the pure gravitational growth regime.

The $322\rightarrow211$ mode achieves a maximum characteristic strain of around $10^{-22}$ for $m_b \sim 10^{-12}$ eV/c$^2$ and GeV/$f \sim 5\times 10^{-19}$. As this maximal transition signal occurs in the moderate self-interaction regime, and not in the short-lived transitional zone in the gravitational regime, such signals are long-lived and have potential observational prospects. This is opposed to non-self-interacting models where, due to their short observable periods, transition signals generally do not have observational prospects. 
When only the 211 state is occupied, the 211 annihilation signal is dominant, peaking at $h \sim 10^{-23}$ for $m_b \sim 10^{-12}$ eV/c$^2$.
 Finally, 322 annihilation is only appreciable when the 322 state is occupied with bosons at higher $\alpha$ values. 

The emission frequencies for the 211 and 322 annihilations are broadly similar for a given $m_b$, whereas the transition signal is at a much lower frequency. For annihilation signals, the GW frequency corresponds to twice the energy of the annihilated particles, whereas for transition signals, it is the difference between the energy levels. The energy of the 211 and 322 levels differs only to the second order in $\alpha$ [see Equation~(\ref{eq:Energy Equation})], and thus for small $\alpha$ as considered here, the energy of the 211 and 322 levels differ by a small relative amount. This leads to similar annihilation frequencies but a much lower transition energy difference than the total particle energy, leading to low-frequency transition GWs.

If we assume an older age of $1.0\times10^6$ yr for Cygnus X-1, the results are broadly similar. However, as there has been more time for the cloud to decay, the strength of the GW signal is generally weaker. The results for MOA-2011 are also similar to those in Figure~\ref{fig:Cygnus 1e5 yr strain}. However, because the mass of the BH is smaller and subsequently MOA-2011 more strongly couples to heavier bosons, the peak GW strain is shifted to a higher boson mass. See details in \ref{sec:AdditionalResults}.

\subsection{Gravitational-wave frequency drift}
\label{subsec:fdot}

\begin{figure}
    \centering
    \includegraphics[width=1.0\textwidth]{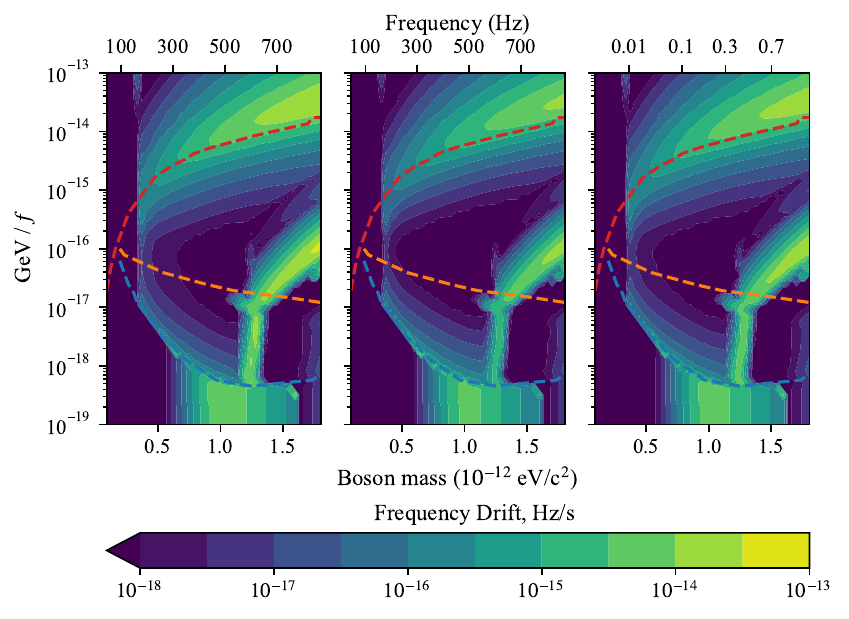}
    \caption{GW frequency drift, $\dot{\nu}$, for signals from Cygnus X-1, assuming parameters as in Table~\ref{tab:Cygnus parameters} and an age of $1.0\times10^5$ yr. The left, middle, and right panels show $\dot{\nu}$ for $211$ annihilation, $322$ annihilation, and $322\rightarrow211$ transition signals, respectively, as a function of boson mass and interaction parameter. Dashed curves divide the parameter space into four regimes, from top to bottom: the no spin-down regime, strong self-interaction regime, moderate self-interaction regime, and gravitational regime.}
    \label{fig:Cygnus 1e5 yr drift}
\end{figure}

The GW signals produced by ultralight boson clouds are expected to be quasi-monochromatic, with a small frequency drift due to BH mass evolution, the boson cloud evolution, higher-order corrections to the energy levels, etc.~\cite{Baryakhtar2021,Isi2019}.
Desired search sensitivities may not be achieved when the signal frequency evolution is faster than what can be covered by the search configuration. 
In Figure~\ref{fig:Cygnus 1e5 yr drift}, we show in coloured contours the signal frequency drift (i.e., the first time derivative of the signal frequency, $\dot{\nu}$) for the three GW emission mechanisms. From left to right, we show $\dot{\nu}$ for the 211 annihilation, 322 annihilation, and $322\rightarrow 211$ transition signals, respectively. We again assume the signals come from Cygnus X-1 at an age of $10^5$~yr.

For the moderate and strong self-interaction regimes, the frequency drift is generally lower than the frequency drift expected within the gravitational regime. The notable exception to this is along the narrow bright region extending from the boundary of the gravitational regime at $m_b \sim 1.2\times10^{12}$ eV/$c^2$ to the maximal boson mass considered within the strong self-interaction regime. This narrow region corresponds to the cloud's evolution stage when the 211 level has decayed sufficiently, and the 322 level becomes dominant. 
The frequency drift is also larger ($\sim 10^{-14}$--$10^{-13}$ Hz/s) at the boundary between the strong self-interaction and no-spin-down regimes. 
However, for the entire parameter space considered, the frequency drift is at most comparable to $10^{-14}$--$10^{-13}$ Hz/s, lower than the maximum $\dot{\nu}$ covered by the search conducted by Sun \textit{et al.}~\cite{Sun2020}. Thus, the effects of the frequency drift for the signal model considered in this paper do not impact the sensitivity achieved in the past search, and the existing results can be interpreted using the model involving self-interaction.

\section{Observational prospects}
\label{sec:results}
In this section, we estimate the search sensitivity that current and future detectors can achieve targeting known isolated BHs or BHs in binary systems (section \ref{sec:sensitivity}). We update the constraints from a past search targeting Cygnus X-1 and project the ultralight scalar parameter space reachable by future detectors, taking Cygnus X-1 and MOA-2011 as examples (section \ref{sec:observable parameter space}). 
\subsection{Projected search sensitivity with current and future detectors}
\label{sec:sensitivity}
Determining the GW search sensitivity from first principles is a non-trivial task. In general, it requires the characterization of the recovery rate of synthetic signals injected into real or simulated data. The recovery rate depends on aspects of the search methodology, such as the algorithm, integration time, template bank, properties of the BH (intrinsic properties, mass and spin; extrinsic properties, orientation and sky position) and uncertainties thereof, and features of the expected signal, such as the signal frequency, frequency evolution, and polarization of the signal~\cite{Isi2019,Riles2023}. Treating these complexities empirically, e.g. using a Monte-Carlo methodology, is possible, but the required computational cost can be high~\cite{Isi2019}.
In this study, instead of taking an empirical approach, we utilize existing search sensitivities for GWs from scalar clouds (without incomplete treatments of boson self-interactions), as presented in Refs.~\cite{Sun2020,Isi2019}, to analytically derive the estimated search sensitivities across the parameter space (boson mass and self-interaction parameter) for current and next-generation detectors.
These existing sensitivity estimates are obtained using a particular methodology, the hidden-Markov-model-based semi-coherent search algorithm~\cite{Suvorova2016,Sun2018,Isi2019}, but generally apply to alternative semi-coherent search methods. The sensitivities in existing literature are usually reported in terms of a characteristic strain, $h_0$, for the 211 annihilation mode, which can be converted to the effective strain (the strain an observer on Earth would measure depending on the orientation of the BH-boson system) for every mode we consider in this study. See detailed conversion in \ref{sec:Orientation Discussion}.

The analytical scaling for the minimum $h_0$ that is detectable 95\% of the time (i.e. 95\% confidence level), denoted by $h_0^{95\%}$, at a signal frequency, $\nu$, is given by~\cite{Sun2020,Isi2019}:
\begin{equation}
    h_0^{95\%}(\nu)\propto S_h^{1/2}(\nu)(T_{\rm coh} T_{\rm obs})^{-1/4},
    \label{eq:scaling}
\end{equation}
where $S_h^{1/2}(\nu)$ is the detector noise amplitude spectral density (ASD) at $\nu$, $T_{\rm obs}$ is the total observing time of the search, and $T_{\rm coh}$ the coherent integration time (i.e., $T_{\rm obs}$ is divided into multiple $T_{\rm coh}$ segments, and the summation across segments is incoherent). 
For a fixed total observing time, search sensitivity improves for longer $T_{\rm coh}$. 
The coherent time $T_{\rm coh}$ can be regarded as the time scale for the signal frequency to drift out of a discrete frequency bin considered in the hidden Markov model, beyond which the signal power is leaked into adjacent bins. Thus, the choice of $T_{\rm coh}$ is limited by the frequency drift rate of the signal (also see Section~\ref{subsec:fdot}).

\begin{figure}
    \centering
    \includegraphics[width=1.0\textwidth]{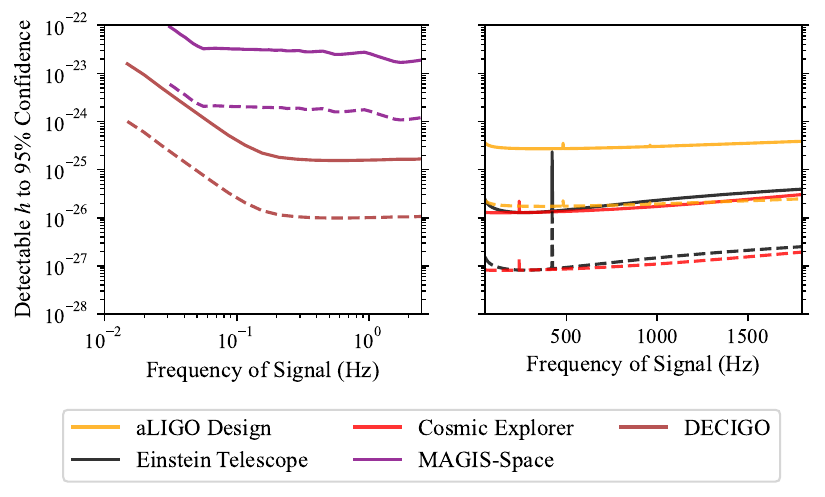}
    \caption{Projected search sensitivity on the effective GW strain with current and next-generation GW detectors (assuming a single detector of each type) to 322$\rightarrow$211 transition (left) and 211 and 322 annihilation signals (right) from Cygnus X-1 (solid lines) and MOA-2011 (dashed lines). The total observing time for both BHs is set to one year. Coherent integration time is set to 10 days for both BHs.}
    \label{fig:Combined sensitivity}
\end{figure}

In Figure \ref{fig:Combined sensitivity}, we show the projected search sensitivity to the effective GW strain as a function of the signal frequency for current and next-generation detectors. 
The left and right panels are for $322\rightarrow211$ transition, and $211$ and $322$ annihilation signals, respectively.
Solid and dashed curves correspond to estimated sensitivities to signals from a BH in a binary system like Cygnus X-1, and an isolated BH like MOA-2011, respectively, and under the assumption of high initial spin. 
We assume a total observing time of $T_{\rm obs}=1$ yr and a coherent integration time of $T_{\rm coh}=10$ d and analytically calculate the estimated sensitivity using existing results in Refs.~\cite{Sun2020,Isi2019} and the scaling in Equation~
(\ref{eq:scaling}).\footnote{For Cygnus X-1, we refer to the search sensitivity achieved in the O2 search~\cite{Sun2020} in the band of 255--256~Hz. Note that there is a slight frequency-dependent sensitivity loss at higher frequencies presented in Ref.~\cite{Sun2020}, due to the fact that the incoherent summation of the orbital sideband powers in a binary system leads to larger sensitivity loss at higher frequencies~\cite{Suvorova2016}. This effect can be surmounted with improved search methods, e.g., \cite{Suvorova2017}, when better orbit measurements are available. Subsequently, the frequency-dependent sensitivity loss is not considered in this paper.}
Since the searches targeting isolated BHs are not affected by the orbital motion, the sensitivity for isolated BHs is generally a factor of 5--8 times better than that for BHs in binaries with the search technique and configuration assumed in this study. 
In addition, the projected sensitivity for Cygnus X-1 is obtained using the real O2 search results, where the non-Gaussain, non-stationary detector noise and operation duty cycles lead to a factor of $\sim 2$ sensitivity loss compared to sensitivities in ideal Gaussian-noise simulations. Thus, overall, the estimates shown in dashed curves for MOA-2011 (based on simulations for isolated systems) are about an order of magnitude better than those in solid curves for Cygnus X-1 (based on real search upper limits for binary systems). 

As expected, the search sensitivity with next-generation ground-based detectors (Cosmic Explorer and Einstein Telescope) is improved over current-generation detectors (aLIGO Design~\cite{Aasi_2015,asdcurves})\footnote{Further upgrading stages~\cite{ligo_posto5_report,Abbott-LRR} within the existing infrastructure after aLIGO achieves the design sensitivity are not included.} by more than an order of magnitude. The future space-based detectors (MAGIS-Space and DECIGO) would cover a much lower frequency range and have sensitivity to transition signals.

\subsection{Observable parameter space}
\label{sec:observable parameter space}
We continue to take Cygnus X-1 with an assumed age of $1.0\times10^5$ yr as an example and demonstrate the observable parameter space of the ultralight scalars. 

\begin{figure}
    \centering
    \includegraphics[width=1.0\textwidth]{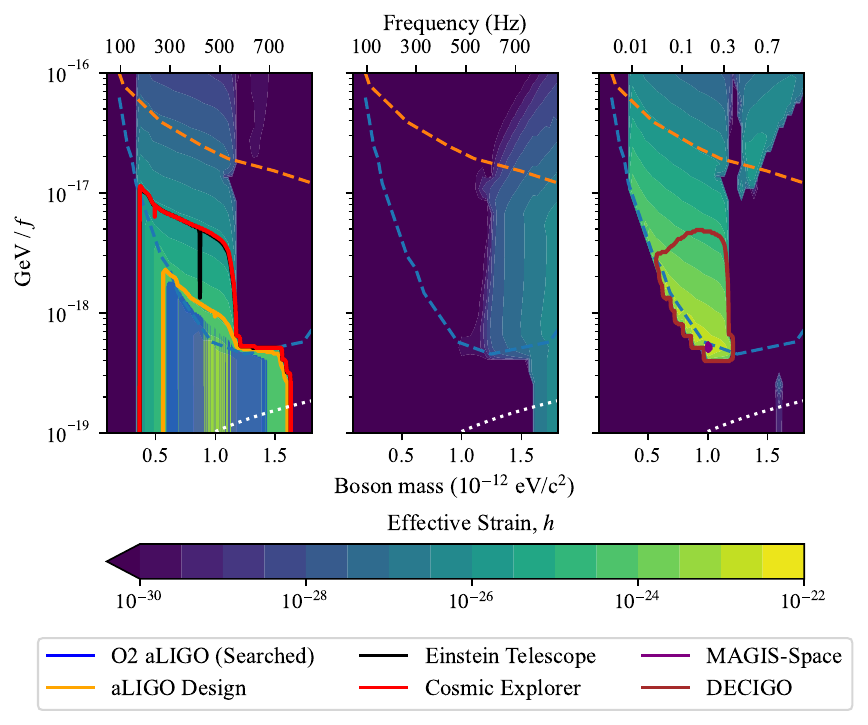}
    \caption{Parameter space reachable by current and next-generation detectors (enclosed by the solid curves; see legend) for (left) 211 annihilation, (middle) 322 annihilation, and (right) $322\rightarrow211$ transition signals from an ultralight scalar cloud around Cygnus X-1, assuming parameters as in Table~\ref{tab:Cygnus parameters} and an age of $1.0\times10^5$ yr. The dashed curves divide the parameter spaces into three regimes: (top) the strong self-interaction regime, (middle) the moderate self-interaction regime, and (bottom) the gravitational regime. The reachable regions for Einstein Telescope (enclosed by black curve) and Cosmic Explorer (enclosed by red curve) largely overlap. The parameter space reachable by MAGIS-Space is limited to the brightest spot in the right panel in this example. In other cases it can probe a larger parameter space; see for, e.g., MOA-2011 in Figure~\ref{fig:MOAAgeSearch}. For reference, the QCD axion's mass to $1/f$ relation is shown by a dotted white line (in the bottom right corner of each plot). Axion interactions with Standard Model photons, electrons, and nuclei result in exclusions in typical models of GeV/$f\gtrsim 10^{-8}-10^{-9}$\cite{Workman:2022ynf}, complementary to the region shown here. See also~\cite{Baryakhtar2021,Hoof:2024quk} for constraints from BH spin measurements.}
    \label{fig:Probeable Space Cygnus 1e5}
\end{figure}

In Figure~\ref{fig:Probeable Space Cygnus 1e5}, we compare the search sensitivity presented in Figure~\ref{fig:Combined sensitivity} to the GW strain shown in Figure~\ref{fig:Cygnus 1e5 yr strain}. We find that for particular parameter space, the 211 annihilation signals can be reached by current and next-generation detectors, and the $322\rightarrow211$ transition signals can be potentially observable with the proposed space-based deci-hertz detectors. The 322 annihilation signals, by contrast, would be too weak to be detectable by either current or next-generation detectors. 

The coloured contours in Figure~\ref{fig:Probeable Space Cygnus 1e5} are the same as those in Figure~\ref{fig:Cygnus 1e5 yr strain}, with the potentially observable parameter space for 211 annihilation signals (left panel) and for $322\rightarrow211$ transition signals (right panel) enclosed by the solid thick curves. See the legend for the assumed detectors. 
The updated exclusion of the parameter space (95~\% confidence level), given the assumed BH parameters,  converted from the existing limits set by the past search~\cite{Sun2020} using the new self-interaction model is shaded in blue. The gaps in the shaded area are noise-contaminated frequency bands where no constraints can be placed.
Note that the reachable regions for Einstein Telescope (enclosed by black curve) and Cosmic Explorer (enclosed by red curve) largely overlap.  
As before, the orange and blue dashed curves divide the parameter space into the strong self-interaction regime, moderate self-interaction regime, and gravitational regime, from top to bottom. We omit the no spin-down regime in this figure since signals in that parameter space are beyond the detection capability of all detectors considered here.
 
We derive a disfavored mass range of approximately $[0.6,1.4]\times 10^{-12}$ eV/c$^2$ (mostly in the gravitational regime), by directly converting the strain upper limits obtained in the aLIGO O2 search targeting Cygnus X-1~\cite{Sun2020}.
The results are broadly consistent with those presented in Ref.~\cite{Sun2020}, where a disfavoured mass range of $[0.63,1.32]\times10^{-12}$~eV/$c^2$ is derived without considering self-interactions (assuming Cygnus X-1 is $1.0\times10^5$ yr old). 
When considering non-negligible boson self-interaction, the past study in Ref.~\cite{Sun2020} used an old `bosenova' model~\cite{Yoshino2015} to calculate the expected signal strain, and a mass range of $[0.96,1.55]\times10^{-12}$ eV/$c^2$ was excluded for $f \sim 10^{15}$ GeV.
However, we do not find that ultralight scalars with a self-interaction parameter $f \sim 10^{15}$ GeV (within the strong self-interaction regime) could be detected using the updated self-interaction model~\cite{Baryakhtar2021}. Hence, no boson mass range can be reached in the strong self-interaction regime (which includes the parameter space relevant to axion-like particle dark matter) based on the aLIGO O2 search results targeting Cygnus X-1. 

We see that the prospects of future searches for ultralight scalars around BHs in binaries such as Cygnus X-1 at current-generation ground-based detectors (aLIGO at design sensitivity) are largely limited to probing 211 annihilation signals in the gravitational regime and a small portion of the adjacent moderate self-interaction regime with $f \sim 10^{18}$~GeV. 
Next-generation ground-based detectors (Einstein Telescope and Cosmic Explorer) extend the reach of 211 annihilation signals to a larger region of the moderate self-interaction regime with $f \sim 10^{17}$~GeV, but cannot reach into the strong self-interaction regime.  
For the proposed deci-hertz space-based detectors (DECIGO and MAGIS-Space), we see that the $322 \rightarrow 211$ transition signals in part of the moderate self-interaction regime fall within the sensitive range of the detectors. The conclusions are broadly similar for an older Cygnus X-1 assumption and for MOA-2011, with slightly improved reach for the isolated system and overall worse sensitivity for older systems. See \ref{sec:AdditionalResults} for further details corresponding to Cygnus X-1 with an age of $10^6$~yr and MOA-2011. 

Considering the overlapping parameter space regions enclosed in the left and right panels of Figures~\ref{fig:Probeable Space Cygnus 1e5}, one particularly interesting possibility with next-generation ground-based and deci-hertz space-based detectors is to search for a multi-band signal from the same BH-boson system. The relative difference in signal strength would break the degeneracy between the orientation of the BH-boson system and the strain, and could in principle allow more accurate characterisation of not just the mass, but also the self-interaction parameter of the bosons. Moreover, a detection in two different frequency bands would strongly confirm the existence of ultralight bosons, as opposed to other possible sources of noise or other types of continuous GWs.


\section{Conclusions}
\label{sec:conclusion}

In this paper, we use the improved treatment of particle self-interactions in superradiance~\cite{Baryakhtar2021} to investigate the GW emission from ultralight superradiant scalars in order to understand the prospects for detection at current and future observatories in the mass-interaction parameter space. We update the results of a past search~\cite{Sun2020} for ultralight scalars around Cygnus X-1, assuming high initial spin and final spin determined by superradiant evolution. In the gravitational regime where particle self-interactions are negligible, the results disfavour scalars with masses in a range of [$0.6,1.4$]$\times 10^{-12}$ eV/c$^2$ ([$0.6,1.1$]$\times 10^{-12}$ eV/c$^2$), for Cygnus X-1 with an assumed age of $1.0\times10^{5}$ years ($1.0\times 10^6$ years), broadly consistent with the results of the past search. However, we do not find evidence to disfavor ultralight scalars with a self-interaction parameter of $\sim 10^{15}$ GeV, as opposed to the results from using the previous `bosenova' model where the energy exchange process between different bound states is not adequately considered. 

We then estimate the parameter space of ultralight scalars reachable with current and next-generation detectors by studying two specific galactic BHs as examples, Cygnus X-1 (a BH in a binary) and MOA-2011 (an isolated BH). Similar nearby BHs in our galaxy with well-measured mass and spin (and orbital parameters if the BH is in a binary) are interesting targets for future GW searches. With next-generation terrestrial GW detectors, we demonstrate that the accessible parameter space of ultralight scalars around these BHs can be extended into the moderate self-interacting regime, improving by a factor of two to five in interaction parameter $f$ compared to current-generation detectors. We will be able to interrogate the existence of ultralight scalars in the mass range of $\sim 3\times10^{-13}$--$3\times10^{-12}$ eV/c$^2$ across the gravitational and moderate self-interaction regimes by targeting galactic BHs like Cygnus X-1 and MOA-2011.

We note that in this work we have only considered central measured values on some BH parameters and assumptions on others (like the initial spin and age) that have large uncertainties, so our projections on the reachable parameter space cannot be considered as true exclusions in the scalar particle parameter space. Data-driven constraints obtained from GW searches marginalized over the BH uncertainties would be of great interest. In particular, more accurate and precise measurements of the BH properties, e.g., mass, spin, age, and orbital parameters (if relevant), with future advanced telescopes will lead to smaller uncertainties in the estimates of the signal strength. By comparing the search sensitivities (with particular false alarm and false dismissal probabilities) to the estimated signals by taking into consideration the signal uncertainties propagated from the BH properties, one can obtain robust constraints on the existence of ultralight scalars.

Unlike the case of gravitational superradiance, in the presence of strong self-interactions, high-spin, old BHs may not directly exclude scalars in the corresponding mass range since strong self-interactions can halt BH spin-down.
On the other hand, we do not find observational GW prospects in the strong self-interaction and no spin-down regimes: the particle self-interactions suppress the growth of superradiant clouds significantly and hence the GW emission becomes too weak to be observed even with next-generation detectors. In particular, observable signals would be in conflict with the high, present-day estimated spin of Cygnus X-1.
In general, preferred GW search targets in our galaxy would be BHs with moderate or low present-day spins (such that observable clouds may exist), whose masses, spins, ages, orientations, and orbital parameters (where applicable) are well-measured. 

There are many future directions to explore self-interacting scalar superradiance. Constraints in the mass-interaction plane using BH spin measurements from X-ray binary systems have been placed~\cite{Baryakhtar2021}, and it would be interesting to also take into account self-interactions when analysing BH binary catalogs observed by the LIGO-Virgo-KAGRA network~\cite{Ng:2019jsx,Ng2021}.

This analysis has been limited to a relatively low boson mass regime, with gravitational fine structure constant $\alpha\leq0.2$. Here, the superradiance cloud can be exactly described by a two-level system of 211 and 322. Further numerical work is required to extend the understanding of self-interactions to a higher $\alpha$ regime, as both the approximations used in our analysis break down and the two-level system fails to capture the full dynamics of the cloud for $\alpha\gtrsim 0.3$. A full relativistic analysis for scalars, such as was done for vectors in Ref.~\cite{East:2017mrj}, is very challenging, however, due to the longer superradiance times and resulting large dynamical range of the scalar system.

Multi-band searches for ultralight scalars, targeting component BHs in inspirals in ground-based observatories by using the inspiral signals observed in the space-based detector band, have been discussed~\cite{Ng:2020jqd}. The presence of self-interactions opens up another interesting multi-band possibility: while scalar transition GW signals do not have promising observational prospects when particle self-interaction is negligible, they are potentially observable in the moderate self-interaction regime with future deci-hertz space-based detectors. Joint observation of annihilation and transition signals would provide clear evidence to identify the signal as originating from scalar superradiance rather than other possible continuous wave sources, and to determine the particle mass and self-interaction scale.

\section*{Acknowledgments}
The authors thank Nils Siemonsen and William East for the discussion about rates in \texttt{SuperRad}.
Numerical simulations in this paper were performed on the OzSTAR national facility at Swinburne University of Technology. The OzSTAR program receives funding in part from the Astronomy National Collaborative Research Infrastructure Strategy (NCRIS) allocation provided by the Australian Government, and from the Victorian Higher Education State Investment Fund (VHESIF) provided by the Victorian Government.
This research is supported by the Australian Research Council Centre of Excellence for Gravitational Wave Discovery (OzGrav), Project Numbers CE170100004 and CE230100016. L.S. is also supported by the Australian Research Council Discovery Early Career Researcher Award, Project Number DE240100206. Research at Perimeter Institute is supported in part by the Government of Canada through the Department of Innovation, Science and Economic Development and by the Province of Ontario through the Ministry of Colleges and Universities. M.B.\ is supported by the U.S. Department of Energy Office of Science under Award Number DE-SC0024375 and the Department of Physics and College of Arts and Science at the University of Washington.

\appendix
\addtocontents{toc}{\fixappendix}
\section{Angular dependence of gravitational-wave strain}
\label{sec:Orientation Discussion}

The GWs emitted from ultralight boson clouds are not isotropic and, in general, have an angular dependence, i.e., the GW power (and thus strain) expected from the cloud depends on its inclination to an observer~\cite{Isi2019,Siemonsen2023}. Subsequently, when determining whether a given GW detector could detect signals from the cloud, it is necessary to take the system orientation into account.

In this paper, we determine the angular dependence of the GW power for each of the three GW emission modes of interest and calculate the \emph{effective} GW strain, $h$, an observer would measure with a given inclination of the BH-boson system. We begin with the characteristic strain ($h_0$), an angle-independent quantity defined for a given mode as~\cite{Isi2019}:
\begin{equation}
    h_{0, \rm mode} =\left(\frac{10G P_{\rm mode}}{c^3 r^2 \omega_{\rm mode}^2}\right)^{1/2},
\end{equation}
with $P_{\rm mode}$ the GW emission power in that mode, and $\omega_{\rm mode}$ the angular frequency of that mode's GW signal.

Recall that an inclination of $0^{\circ}$ ($180^{\circ}$) corresponds to a BH spin pointed directly at (away from) Earth; this is a `face-on' (`face-off') orientation. For a face-on/face-off system emitting GWs in the 211 annihilation mode, the effective GW strain, $h$, is equal to the characteristic strain, $h_0$.
For a non face-on/face-off system, or for the 322 annihilation and $322\rightarrow211$ transition modes, the effective strain is not equal to the characteristic strain. However, we can calculate each mode's characteristic strain from the emission power, and then derive the angle-dependent effective strain for a particular system orientation based on the angular distribution of each mode. In particular, the effective strain at a given inclination, $\iota$, is given by:
\begin{equation}
    h_{\rm mode}(\iota)= \sqrt{\frac{8 \pi}{5}} \, h_{0, \rm mode} \left[\frac{\left(\frac{dP}{d\Omega}\right)_{\rm mode}\left(\iota \right)}{P_{\rm mode}} \right]^{1/2}.
    \label{eq:angular dependence}
\end{equation}
The constant $\sqrt{\frac{8 \pi}{5}}=\left[\frac{{P_{\rm 211 \, An.}}}{{\left(\frac{dP}{d\Omega}\right)_{\rm 211 \, An.}\left(\iota = 0 \right)}}\right]^{1/2}$ relates the characteristic strain to the emission power for an arbitrary mode, using the relation derived in the 211 annihilation mode.

For 211 annihilation signals, the angular dependence of the GW power to the leading order of $\alpha$ is given  by \cite{Arvanitaki2011}:
\begin{equation}
    \left(\frac{dP}{d\Omega}\right)_{\rm 211 \, An.} \propto 35 + 28 {\rm cos}2\iota + {\rm cos}4\iota .
    \label{eq:211 angular dependence}
\end{equation}
For $322\rightarrow211$ transitions, the angular dependence is \cite{Baryakhtar2021}:
\begin{equation}
    \left(\frac{dP}{d\Omega}\right)_{\rm Tr.} \propto  \frac{2^5}{3^6 5^8}\left(1-{\rm cos}^4 \iota \right) + \frac{\left(27 + 28 {\rm cos} 2 \iota + 9 {\rm cos} 4 \iota \right){\rm sin}^2 \iota}{2^2 3^6 5^{10} 7^2} .
    \label{eq:tr angular dependence}
\end{equation}
For 322 annihilation, we used the numerical superradiance package \texttt{SuperRad} \cite{Siemonsen2023} for the numerically computed angular dependence. An analytical formula can be found in Ref.~\cite{Arvanitaki2015}.

\begin{figure}
    \centering
    \includegraphics[width=1.0\textwidth]{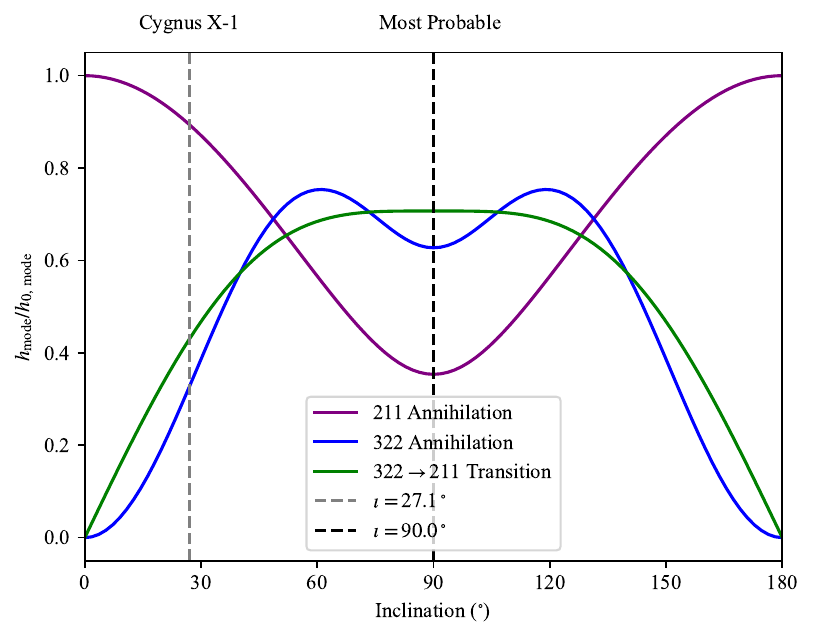}
    \caption{GW strain as a function of the inclination angle of the BH. The GW strain of 211 annihilation, 322 annihilation, and 322$\rightarrow$211 transition signals is shown relative to $h_0$ in each mode. The inclination of Cygnus X-1 and the most probable edge-on inclination are marked by vertical dashed lines. The edge-on inclination ($\iota = 90^\circ$) is used for the estimates for MOA-2011 in this study.}
    \label{fig:OrientationCorrection}
\end{figure}

In Figure \ref{fig:OrientationCorrection}, we utilise Equation~(\ref{eq:angular dependence}) to show the inclination dependence of the GW strain for 211 annihilation, 322 annihilation, and $322\rightarrow211$ transition signals. An inclination of $90^{\circ}$ corresponds to a BH spin perpendicular to the direction pointing towards Earth; this is an `edge-on' orientation. Accounting for the distribution of BH spin orientations, edge-on is the most likely orientation, and face-on/face-off is the least likely.

It can be seen that the maximal strain for 211 annihilation occurs for a perfectly face-on/face-off system and falls to the minimum of approximately 0.353 times the characteristic strain for an edge-on system. By contrast, the maximal strain for $322\rightarrow211$ transition signals occurs for an edge-on system. The strain of the 322 annihilation signals reaches the maximal value at a near, but not perfectly, edge-on orientation. It is also notable that (to the orders considered in this paper), the 322 annihilation and $322\rightarrow211$ transition signals are expected to have no observable strain for a perfectly face-on/face-off system.

The inclination for Cygnus X-1 is favourable for the 211 annihilation emission but not optimized for the 322 annihilation and $322\rightarrow211$ transition signals. The assumed edge-on orientation for MOA-2011 in this study is favourable for the observation of 322 annihilation and $322\rightarrow211$ transition signals, but correspondingly is the worst-case scenario for 211 annihilation signals.

\section{Retrospective initial mass determination}
\label{sec: Mass Evolution}

As the mass of a BH evolves due to the growth of an ultralight boson cloud, if superradiance happens, and generally only the final mass of a BH can be observed, it is necessary to retrospectively determine the initial BH mass. The standard approach to this problem \cite{Sun2020,Baryakhtar2021} is to assume an initial BH spin, and assume the BH has developed a \textit{single} $m$-level cloud to its maximum size. The BH's initial mass is then related to its final mass and initial spin by an analytical relation. Following the procedure of Ref.~\cite{Baryakhtar2021}, this relation for an arbitrary $m$-level is:
\begin{equation}
    M_i = \frac{\hbar c}{G m_b}\frac{-m \sqrt{4 \alpha ^2+m^2} \sqrt{4 \alpha ^2-4 \alpha  m \chi _i+m^2}+m^3+4 \alpha ^2 m}{2 \left(4 \alpha ^2 \chi _i+m^2 \chi _i\right)}, \label{eq:Initial mass}
\end{equation}
where $M_i$ is the initial BH mass, $\chi_i$ the initial BH spin, and $\alpha$ the gravitational fine-coupling constant at the BH's \textit{final} mass. This approach can also be extended for successively growing levels (e.g., $m=1$ and $m=2$ levels) by applying it in succession.

However, this approach does not work when considering particle self-interactions, as the self-interaction drives the concurrent growth of multiple $m$ levels. Thus, the analytical approach using Equation~(\ref{eq:Initial mass}) is no longer valid, and an alternative approach is needed to estimate the initial BH mass.

\begin{figure}
    \centering
    \includegraphics[width=1.0\textwidth]{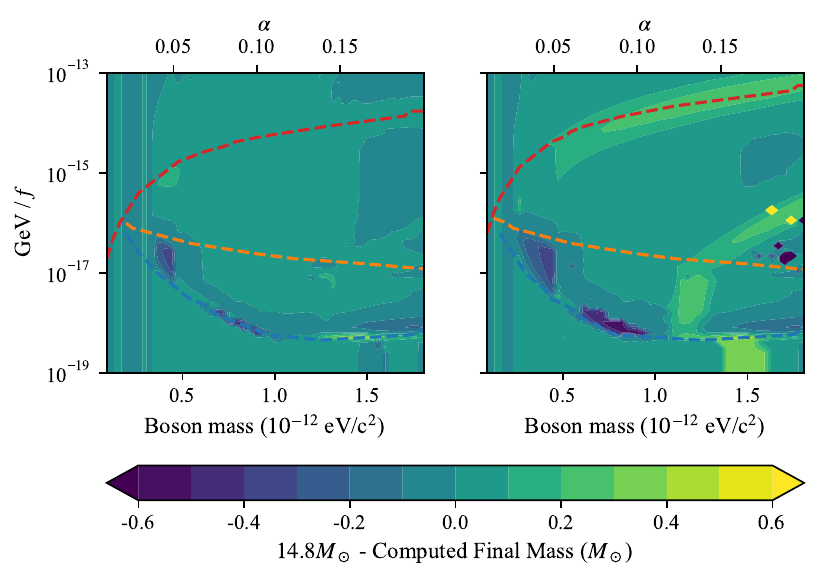}
    \caption{Mass residuals achieved for Cygnus X-1 with an assumed age of $1.0\times 10^5$ yr (left) and $1.0\times 10^6$ yr (right) and other parameters as in Table~\ref{tab:Cygnus parameters}.}
    \label{fig:CygMassResiduals}
\end{figure}

For Cygnus X-1, we adopt an iterative approach. For each BH-boson system (each set of boson mass and interaction parameter), we use the analytical approximation, Equation~(\ref{eq:Initial mass}), to make a guess for the initial BH mass and then solve for the evolution of the BH-boson system to determine the final BH mass. The difference between the final mass achieved in our numerical calculation and the measured mass, $M_{\rm measured}$, is then used to construct a better guess. The process is then repeated with the new guess as the initial BH mass. Mathematically, the improved guesses are defined with the recursive relation between guess number $j$ and improved guess number $j+1$:
\begin{equation}
    M_i(j+1)=M_i(j)-\left[M_f(j)-M_{\rm measured}\right].
\end{equation}
The process is terminated once $|M_f-M_{\rm measured}| < 0.6 M_\odot$ is achieved for the parameter space with GW observational prospects. The residuals achieved following this approach are shown in Figure~\ref{fig:CygMassResiduals}.

For MOA-2011, rather than applying this approach, we instead \textit{neglect} the mass evolution, i.e., we assume $M_f \approx M_i$. 
This assumption allows us to calculate the cloud evolution for MOA-2011, which does not have any compelling estimates of the BH age \cite{Sahu2022}, at an arbitrary BH age without spending a significant computational cost. 
Neglecting mass evolution can result in an error of at most a factor of $\sim 3$ in the expected 211 and 322 occupation levels, and correspondingly in the GW strength~\cite{Baryakhtar2021}.

\section{Additional results for Cygnus X-1 and MOA-2011}
\label{sec:AdditionalResults}
\subsection{Cygnus X-1 with an age of $10^6$ yr}

\begin{figure}
    \centering
    \includegraphics[width=1.0\textwidth]{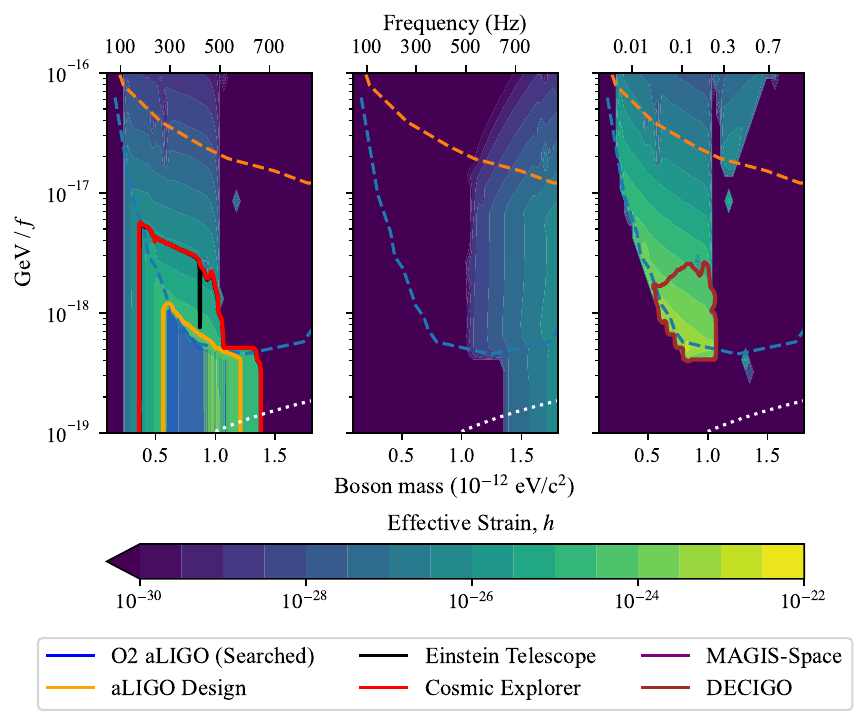}
    \caption{(Similar to Figure~\ref{fig:Probeable Space Cygnus 1e5}) Parameter space reachable by current and next-generation detectors (enclosed by the solid curves; see legend) for signals from Cygnus X-1, assuming parameters as in Table~\ref{tab:Cygnus parameters} and an age of $1.0\times10^6$ yr. The reachable regions for the 211 annihilation, 322 annihilation, and $322\rightarrow211$ transition signals are shown in the left, middle, and right panels, respectively. The dashed curves divide the parameter spaces into three regimes: (top) the strong self-interaction regime, (middle) the moderate self-interaction regime, and (bottom) the gravitational regime. The reachable regions for Einstein Telescope (enclosed by black curve) and Cosmic Explorer (enclosed by red curve) largely overlap. The transition signal is not reachable by MAGIS-Space for this particular example. For reference, the QCD axion's mass to $1/f$ relation is shown by a dotted white line (in the bottom right corner of each plot).}
    \label{fig:Probeable Space Cygnus 1e6}
\end{figure}

In Figure~\ref{fig:Probeable Space Cygnus 1e6}, we show the potentially observable ultralight scalar parameter space for Cygnus X-1, assuming the BH parameters in Table~\ref{tab:Cygnus parameters} and an age of $1.0\times10^6$ yr. This plot is in the same style as Figure~\ref{fig:Probeable Space Cygnus 1e5} and shows the same qualitative morphology (with the same physical explanation). 

Comparing the reachable parameter space between Figures~\ref{fig:Probeable Space Cygnus 1e5} and \ref{fig:Probeable Space Cygnus 1e6}, we note several similarities between the two age assumptions. In both cases, there are observation prospects for 211 annihilation and $322\rightarrow211$ transition, but not for 322 annihilation. 
However, for an older estimated BH age (Figure~\ref{fig:Probeable Space Cygnus 1e6}), the reachable parameter space shrinks. We find that at a BH age of $1.0\times10^6$ yr, the aLIGO O2 search results~\cite{Sun2020} disfavor a mass range of $[0.6,1.1]\times 10^{-12}$ eV/c$^2$, as opposed to the range of $[0.6,1.4]\times 10^{-12}$ eV/c$^2$ found for a BH age of $1.0\times10^5$ yr (Figure~\ref{fig:Probeable Space Cygnus 1e5}).\footnote{An older BH age of $5.0\times10^6$ yr was considered in Ref.~\cite{Sun2020}, leading to a narrower disfavored boson mass range of $[0.6,0.8]\times 10^{-12}$ eV/c$^2$, compared to those presented here. In general, the disfavored boson mass ranges (falling in the gravitational regime) obtained in this study are consistent with those in Ref.~\cite{Sun2020}.} Similarly, with next-generation detectors, the 211 annihilation signals are only expected to be observable up to a boson mass of $\sim 1.4\times 10^{-12}$ eV/c$^2$ at a BH age of $1.0\times10^6$ yr, as opposed to $\sim 1.6\times 10^{-12}$ eV/c$^2$ at a BH age of $1.0\times10^5$ yr. This difference is because, at an older age, the 211 and 322 levels have had a longer time depleting themselves through GW emission and self-interaction, leading to overall lower occupancy of each level and, thus, weaker GW signal strength. 

\subsection{MOA-2011}

\begin{figure}
    \centering
    \includegraphics[width=1.0\textwidth]{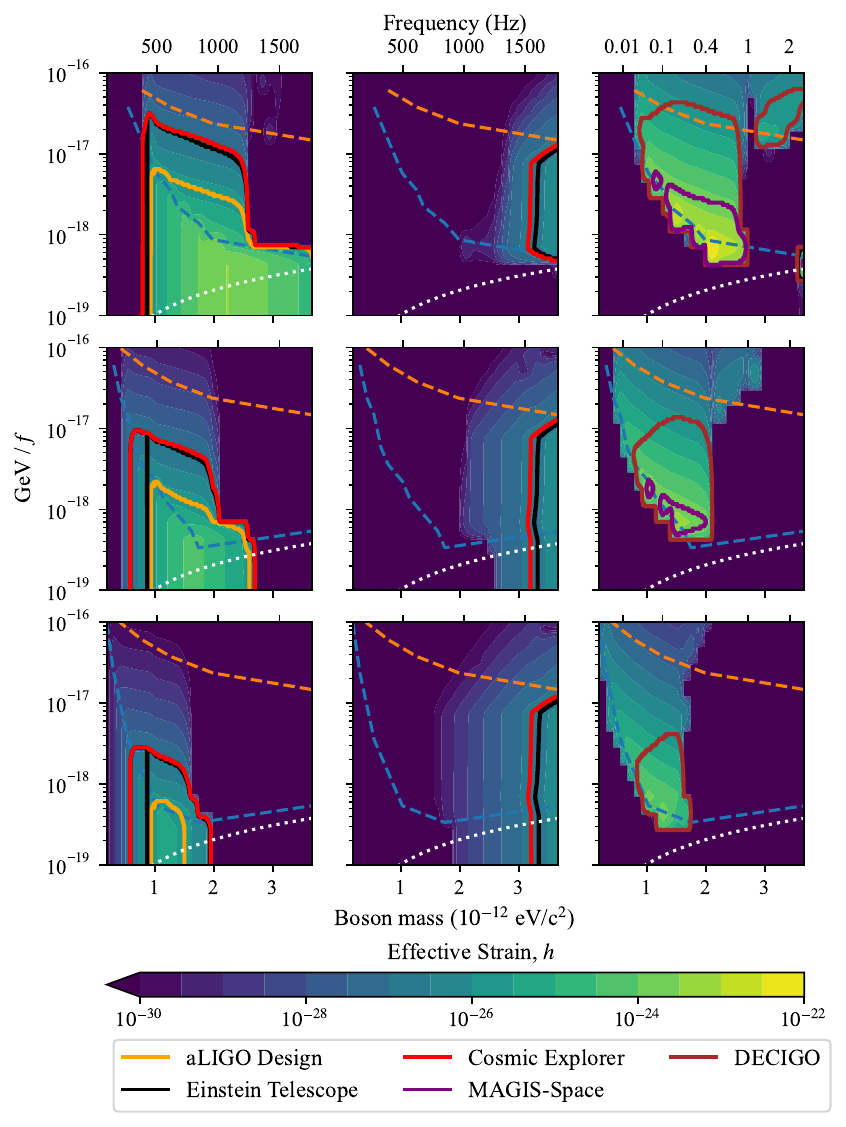}
    \caption{Parameter space reachable by current and next-generation detectors (enclosed by the solid curves; see legend) for signals from MOA-2011 assuming parameters as in Table~\ref{tab:MOA parameters}. Cases are shown for an assumed BH age of $1.0\times10^4$ yr (top row), $1.0\times10^6$ yr (middle row), and $1.0\times 10^8$ yr (bottom row). The reachable regions for 211 annihilation, 322 annihilation, and $322\rightarrow211$ transition signals are shown in the left, middle, and right columns, respectively. The dashed curves divide the parameter spaces into three regimes: (top) the strong self-interaction regime, (middle) the moderate self-interaction regime, and (bottom) the gravitational regime. The reachable regions for Einstein Telescope (enclosed by black curve) and Cosmic Explorer (enclosed by red curve) largely overlap. For reference, the QCD axion's mass to $1/f$ relation is shown by a dotted white line (in the bottom right corner of each plot).}
    \label{fig:MOAAgeSearch}
\end{figure}

In Figure~\ref{fig:MOAAgeSearch}, we show the potentially observable ultralight scalar parameter space for MOA-2011 considering the BH parameters in Table~\ref{tab:MOA parameters} and three assumed BH ages of $1.0\times 10^4$ yr (top), $1.0\times 10^6$ yr (middle), and $1.0\times 10^8$ yr (bottom). For each age assumption (each row in the figure), the GW strain and excluded regions are shown in the same style as Figures~\ref{fig:Probeable Space Cygnus 1e5} and \ref{fig:Probeable Space Cygnus 1e6}.

The qualitative features in the GW strain (e.g., the 211 annihilation dominant region in the gravitational regime) are similar for all ages, as well as to those observed in Figures~\ref{fig:Probeable Space Cygnus 1e5} and \ref{fig:Probeable Space Cygnus 1e6} for Cygnus X-1. 
However, compared to Cygnus X-1, these bright regions occur at higher boson masses as, in general, lighter BHs couple to heavier ultralight bosons. The size and position of the bright regions vary markedly between ages. 
In particular, the size of the bright regions of 211 annihilation and $322\rightarrow211$ transition is the largest for the youngest age assumption and decreases for older age assumptions.
Conversely, the size of the bright region of 322 annihilation increases with the assumed age. This reflects the fact that the 211 level grows first but then depletes itself and is eventually entirely supplanted by the 322 level. As the $322\rightarrow211$ transition requires the occupation of \textit{both} the 211 and 322 levels, it follows the same trend as 211 annihilation.

Considering the reachable parameter space for MOA-2011, one major difference compared to the Cygnus X-1 results is that 322 annihilation signals, though expected to be the weakest among the three modes, are potentially reachable by the next-generation ground-based detectors.
This difference is due to three factors. First, on the technical aspect, better sensitivity can be achieved when targeting an isolated BH like MOA-2011 (see Section~\ref{sec:sensitivity}). Second, the measured distance to MOA-2011 is less than that to Cygnus X-1. Third, we consider an edge-on orientation for MOA-2011, which is a preferred emission direction for 322 annihilation. 
The reachable parameter space for 322 annihilation is generally toward higher boson masses or, equivalently, a higher $\alpha$ regime.
At such a high $\alpha$ regime, more sophisticated numerical calculations are required in addition to the model presented in Ref.~\cite{Baryakhtar2021}. Future improvement in numerical studies will allow us to explore regimes with $\alpha \gtrsim 0.2$, which may lead to improved observational prospects for 322 annihilation signals.

There is no possibility for a multi-band observation of 322 annihilation with either 211 annihilation or $322\rightarrow211$ transition signals, given that 322 annihilation signal is observable \textit{only} when the 322 level is dominant. However, similar to Cygnus X-1, there remains the possibility of observing both 211 annihilation and $322\rightarrow211$ transition signals concurrently. 

\section{Exceptional regimes}
\label{sec:minor_regimes}

Further to the four major regimes discussed in Section~\ref{sec:regimes} and detailed in Ref.~\cite{Baryakhtar2021}, two exceptional regimes are highlighted here for completeness, termed the immediate (level) switch regime and the harmonic equilibrium regime. The dynamics of the exceptional regimes are driven by the interplay of gravitational and self-interactive processes, and subsequently, they occur in the parameter space where the rates of the gravitational and self-interactive processes are comparable, i.e., within the moderate and strong self-interaction regimes.

\subsection{Immediate switch}

\begin{figure}
    \centering
    \includegraphics[width=1.0\textwidth]{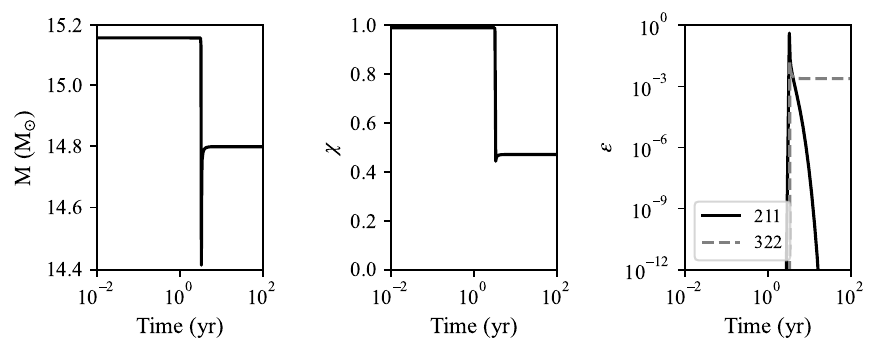}
    \caption{Evolution of a BH-boson system in the immediate (level) switch regime. The left, middle, and right panels show the BH mass, BH spin, and boson cloud occupation numbers, respectively, as a function of time. The ultralight boson has a mass of $m_b=1.5\times10^{-12}$ eV/c$^2$ and a self-interaction parameter of $f=8.4\times10^{16}$~GeV (GeV/$f=1.2\times 10^{-17}$). The associated BH has an initial mass of $M_i = 14.8 M_\odot$ and an initial spin of $\chi_i=0.99$.}
    \label{fig:Axion Burst Regime Evolution}
\end{figure}

A system in the immediate (level) switch regime evolves initially as a normal system in the moderate or strong self-interaction regime; the 211 level grows initially through superradiance, and the 322 level then grows through self-interactive couplings from the 211 to the 322 level. In a normal system, this self-interactive coupling releases some energy in the form of scalar particles emitted from the BH, but then a quasi-equilibrium is achieved. However, in the immediate switch regime, the released energy is sufficiently large to deplete the energy of the system such that a quasi-equilibrium cannot be achieved. Instead, all of the energy in the 211 level is transferred to the 322 level or lost to particle emissions; the dominant level is immediately switched. An example of a system within the immediate switch regime is shown in Figure \ref{fig:Axion Burst Regime Evolution}. In this example, 0.36 $M_\odot$ of energy (constituting 2.3 \%  of the total BH energy) is released in the form of scalar particles emitted to infinity.

At present, the boundaries of the immediate switch regime have not been well studied. However, qualitatively, to drive the immediate level switch, the 211 level must grow to near its maximum size to maximise the energy available for a burst emission of particles, and the self-interaction needs to be sufficiently strong to break the quasi-equilibrium. Thus, the exceptional immediate switch regime occurs near the boundary of the moderate and strong self-interaction regimes.

\subsection{Harmonic equilibrium}
\begin{figure}
    \centering
    \includegraphics[width=.8\textwidth]{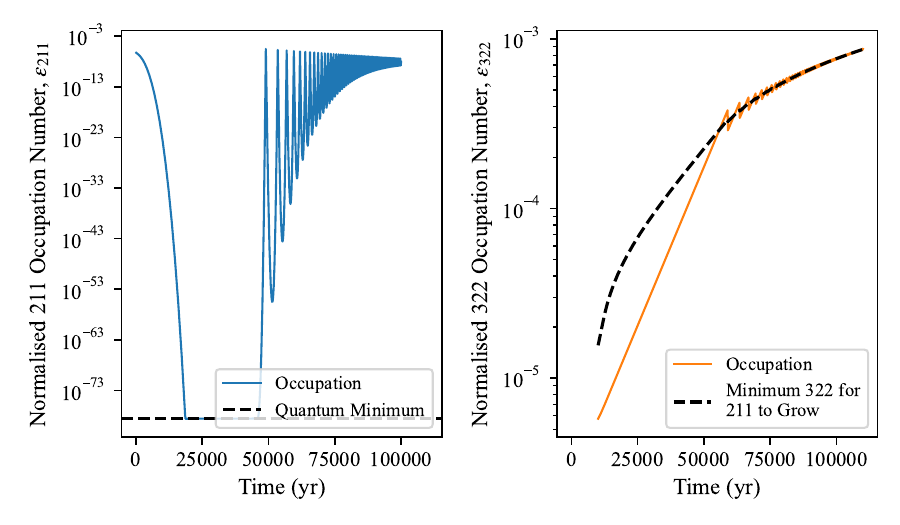}
    \caption{Evolution of a BH-boson system in the harmonic equilibrium regime.
    The left and right panels show the occupation numbers of 211 and 322 levels, respectively, as a function of time. The ultralight boson has a mass of $m_b=1.3\times10^{-12}$ eV/c$^2$ and a self-interaction parameter of $f=3.46\times10^{16}$ GeV. The associated BH has an initial mass of $M_i = 14.8 M_\odot$ and an initial spin of $\chi_i=0.99$.}
    \label{fig:Harmonic Equilibrium Regime Evolution}
\end{figure}

The second exceptional regime is the harmonic equilibrium regime. A system within the harmonic equilibrium regime evolves identically to a normal system within the moderate or strong self-interaction regime until the 211 level decays and the 322 level becomes dominant. However, rather than having the 211 level remain negligible for the remainder of the BH's lifetime, the 211 and 322 levels of the system reach a harmonic equilibrium, with the 211 level alternating between growth and decay. An example is shown in Figure~\ref{fig:Harmonic Equilibrium Regime Evolution}. 
The 211 and 322 level occupations are shown on the left and right, respectively. Note that the vertical axes are different, and the horizontal axes are in a linear timescale to better capture the behavior. 

Starting with the occupation of the $211$ level at its quantum minimum, dynamics in the harmonic equilibrium regime are driven by a cyclic process as follows: 
First, the $322$ level grows until the rate of the $322\times322\rightarrow211\times \rm BH$ process exceeds the absorption rate of the $211$ level by the BH. 
Second, the 211 level grows by the self-interaction process until it is of approximately the same order as the 322 level. 
Third, when the two levels of occupation are comparable, it leads to the depletion of the 322 level until the self-interaction process happens at a slower rate and the 211 level decays again. The cycle then repeats, with the timescale and amplitude of oscillations decreasing with each cycle until an equilibrium is reached, where the growth rate of the 211 level due to the self-interaction exactly matches its own absorption rate by the BH, and the growth rate of the 322 level due to superradiance matches its depletion rate due to self-interaction. Subsequently, this equilibrium can only be maintained as long as the 322 level satisfies the superradiant condition.

The harmonic equilibrium regime occurs on the boundary of the moderate and strong self-interaction regimes, with the exact boundaries and timescales remaining to be determined. 

\section*{References}
\bibliographystyle{iopart-num}
\bibliography{refs}

\end{document}